\documentclass[sigconf,10pt]{acmart}

\usepackage{amsmath,amsfonts}
\usepackage{algorithmic}
\usepackage{graphicx}
\usepackage{textcomp}
\usepackage{xcolor}
\usepackage[acronyms,nonumberlist,nopostdot,nomain,nogroupskip]{glossaries}

\usepackage{ifthen}
\usepackage{xspace}
\usepackage{makecell}
\usepackage{subcaption}
\usepackage{multirow}

\usepackage{diagbox}
\usepackage{orcidlink}

\newboolean{showcomments}
\setboolean{showcomments}{true}
\newcommand{\ma}[1]{\ifthenelse{\boolean{showcomments}}{\textcolor{blue}{\textbf{[MA: #1]}}}{}}
\newcommand{\fr}[1]{\ifthenelse{\boolean{showcomments}}{\textcolor{red}{\textbf{[FR: #1]}}}{}}
\newcommand{\fm}[1]{\ifthenelse{\boolean{showcomments}}{\textcolor{purple}{\textbf{[FM: #1]}}}{}}

\newacronym{pdu}{PDU}{protocol data unit}
\newacronym{md}{MD}{mobile device}
\newacronym{es}{ES}{edge server}
\newacronym{iot}{IoT}{internet of things}
\newacronym{ml}{ML}{machine learning}
\newacronym{vr}{VR}{virtual reality}
\newacronym{ar}{AR}{augmented reality}
\newacronym{dnn}{DNN}{deep neural network}
\newacronym{cv}{CV}{computer vision}
\newacronym{mec}{MEC}{mobile edge computing}
\newacronym{dec}{DEC}{distributed edge computing}
\newacronym{siso}{SISO}{single-input single-output}
\newacronym{mimo}{MIMO}{multiple-input multiple-output}
\newacronym{semcom}{SemCom}{semantic communication}
\newacronym{phy}{PHY}{physical layer}
\newacronym{ndp}{NDP}{null data packet}
\newacronym{cfr}{CFR}{channel frequency response}
\newacronym{sdr}{SDR}{software-defined radio}
\newacronym{i/q}{I/Q}{in-phase/quadrature}
\newacronym{kl}{KL}{Kullback–Leibler}
\newacronym{ofdm}{OFDM}{orthogonal frequency division multiplexing}
\newacronym{fft}{FFT}{fast Fourier transform}
\newacronym{ifft}{IFFT}{inverse fast Fourier transform}
\newacronym{awgn}{AWGN}{additive white Gaussian noise}
\newacronym{snr}{SNR}{signal-to-noise ratio}
\newacronym{ib}{IB}{information bottleneck}
\newacronym{vib}{VIB}{variational information bottleneck}
\newacronym{csi}{CSI}{channel state information}
\newacronym{hrr}{HRR}{holographic reduced representations}
\newacronym{mbat}{MBAT}{matrix binding of additive terms}
\newacronym{kd}{KD}{knowledge distillation}
\newacronym{mse}{MSE}{mean-square error}
\newacronym{ber}{BER}{bit error rate}
\newacronym{map}{MAP}{maximum a posteriori}
\newacronym{jscc}{JSCC}{joint source and channel coding}
\newacronym{los}{LoS}{line-of-sight}
\newacronym{nlos}{NLoS}{non-line-of-sight}
\newacronym{tcp}{TCP}{transmission control protocol}
\newacronym{mtu}{MTU}{maximum transmission unit}
\newacronym{qpsk}{QPSK}{quadratic phase-shift keying}
\newacronym{mmwave}{mmWave}{millimeter-wave}
\newacronym{flop}{FLOP}{floating-point operation}
\newacronym{rf}{RF}{radio frequency}
\newacronym{lra}{LRA}{long range arena}

\AtBeginDocument{%
  }

\setcopyright{acmlicensed}
\copyrightyear{2026}
\acmYear{2026}
\acmDOI{XXXXXXX.XXXXXXX}
\acmConference[Technical Report]{Technical Report}{November, 2025}{Boston, MA, USA}
\acmISBN{978-1-4503-XXXX-X/18/06}

\renewcommand\footnotetextcopyrightpermission[1]{} 
\newcommand{\FW}{Semantic Multiplexing\xspace}

\begin{document}
\title{Semantic Multiplexing}

\author{Mohammad Abdi \orcidlink{0000-0001-6737-2632}, Francesca Meneghello \orcidlink{0000-0002-9905-0360
}, and Francesco Restuccia \orcidlink{0000-0002-4999-4507}\vspace{0.2cm}}
\email{{abdi.mo, fr.meneghello, f.restuccia}@northeastern.edu}
\affiliation{%
  \institution{Northeastern University, United States\vspace{0.3cm}}
  \country{}
}

\renewcommand{\shortauthors}{Abdi et al.}

\begin{abstract}

Mobile devices increasingly require the parallel execution of several computing tasks offloaded at the wireless edge. Existing communication systems only support parallel transmissions at the bit level, which fundamentally limits the number of tasks that can be concurrently processed. To address this bottleneck, this paper introduces the new concept of \textit{Semantic Multiplexing}. Our approach shifts stream multiplexing from \emph{bits} to \emph{tasks} by merging multiple task-related compressed representations into a single \emph{semantic representation}. As such, \FW can multiplex more tasks than the number of physical channels without adding antennas or widening bandwidth by \emph{extending} the effective degrees of freedom at the \emph{semantic} layer, without contradicting Shannon capacity rules. We have prototyped \FW on an experimental testbed with Jetson Orin Nano and millimeter-wave software-defined radios and tested its performance on image classification and sentiment analysis while comparing to several existing baselines in semantic communications. Our experiments demonstrate that \FW allows jointly processing multiple tasks at the semantic level while maintaining sufficient task accuracy. For example, image classification accuracy drops by less than 4\% when increasing from 2 to 8 the number of tasks multiplexed over a 4$\times$4 channel. \FW reduces latency, energy consumption, and communication load respectively by up to 8$\times$, 25$\times$ and 54$\times$ compared to the baselines while keeping comparable performance. We pledge to publicly share the complete software codebase and the collected datasets for reproducibility. \vspace{-0.3cm} 
\end{abstract}

\maketitle

\section{Introduction}\label{sec:intro}

Modern mobile devices need to continuously offload computing tasks (e.g., object detection, semantic segmentation) to edge servers to process high-data-rate multimedia inputs without prematurely exhausting their batteries. Such computing tasks need to be executed with ultra-low latency to avoid user discomfort~\cite{gallagher2018cybersickness}. While \gls{mimo} communication allows simultaneously transmitting (i.e., multiplex) multiple streams of bits in the same time/frequency resources, this does not suffice to provide adequate communication support for emerging applications (e.g., mobile virtual reality). \smallskip

As detailed in Section \ref{sec:related-work}, \textit{existing task-driven compression and communication strategies can only process a \textit{single} task at a time}. As such, they are fundamentally limited by the number of data streams currently supported by the wireless standards such as Wi-Fi and 5G. To make a real-world example, Apple's virtual reality headsets are equipped with 12 cameras, among which 2 cameras are for high resolution video and generate 8K frames at 120~Hz each. This implies that 2 frames have to be compressed, transmitted and processed with an end-to-end delay of around 8 milliseconds without compromising the task accuracy~\cite{AppleVisionPro_ProductPage_2025}. Using H.265/HEVC compression and considering 8-bit color depth, each frame occupies 4~MB, corresponding to a data rate of 4~Gbps. State-of-the-art Wi-Fi 7 networks (802.11be) offer up to 2.875~Gbps per \gls{mimo} stream, meaning that at least two streams should be dedicated to each device, thus making the system not scalable~\cite{geraci2025wi}. 

\vspace{-0.2cm}

\begin{figure}[h]
    \centering
    \includegraphics[width=0.92\linewidth]{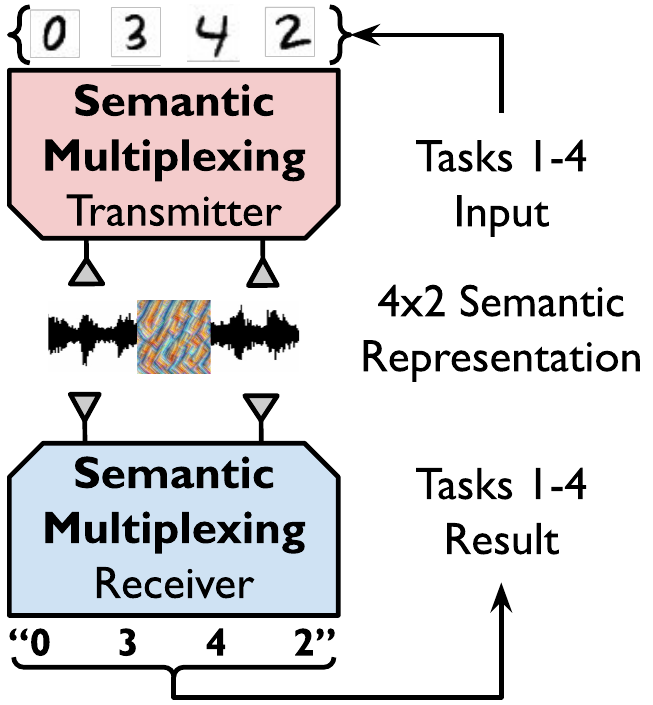}
    \caption{Example of semantic multiplexing four tasks into a single semantic representation spanning two physical streams. At the receiver, the representation is processed to execute the tasks in parallel.\vspace{-0.1cm}}
    \label{fig:overview}
\end{figure}

To address this fundamental challenge, we introduce the new concept of \textit{Semantic Multiplexing}, where we shift stream multiplexing from \emph{bits} to \emph{tasks}.~By merging multiple task-related representations into a single  \emph{semantic representation} that is then multiplexed onto different physical data streams, \FW \emph{extends} the effective degrees of freedom at the \emph{semantic} layer, without adding antennas or widening bandwidth while fully obeying Shannon capacity rules. Figure~\ref{fig:overview} shows an example where four digit recognition tasks (``0'', ``3'', ``4'', and ``2'') are multiplexed and transmitted over two communication channels by semantically superimposing their representations into one physical signal. The latter is then semantically demultiplexed by the receiver to extract in parallel the output of the four tasks \textit{at the same time}.\smallskip

The fundamental technical advance behind \FW is the integration of an additional semantic orthogonalization layer that can be on top of traditional systems such as \gls{mimo}. Critically, \FW's semantically-orthogonal streams can be \textit{simultaneously processed and transmitted} while guaranteeing adequate final task performance for each multiplexed task. By creating \textit{protected channels} for each task, \FW enables both \textit{computation and communication in superposition}, thus drastically reducing the overall task execution time. \smallskip

\noindent \textbf{Technical Challenges.}~The first key challenge is to devise a \textit{task-agnostic} methodology to extract and map several latent representations into physical layer waveforms. For this reason, we propose a new information-theoretical framework where the joint design of communication and computation is enabled by the incorporation of the wireless channel model into the system as a non-learnable function (see Section~\ref{subsec:channel-model}). In stark opposition with prior work that optimizes communication and computing separately \cite{cai2024multi, cai2025end}, our design learns semantic coding in addition to the processing task end-to-end, thus effectively compensating for the distortion patterns caused by the channel while contributing to the processing for task execution. \vspace{0.1cm}

The second challenge is to maintain orthogonality of the learned semantic representations with dynamic channels. As such, we introduce a new channel sounding methodology where pre-defined inputs are sent and processed as \textit{semantic pilots} (see Section~\ref{subsec:adapt}). As the task inference result for such pilots is known at the receiver, the \FW transmitter and receiver are tuned considering the updated wireless channel and back-propagating the task loss through the corresponding computation channel. Prior work \cite{abdi2025phydnns,cai2024multi, cai2025end} only considers communication pilots, thus failing to properly adapt to changing wireless channel conditions as shown in Section~\ref{subsec:task_accuracy}. As \FW is trained in an \textit{end-to-end} manner, the over-the-air channel effects are leveraged as part of the overall computation pipeline, contributing to the task execution.

\smallskip
\noindent \textbf{Summary of Novel Contributions} \smallskip

\noindent $\bullet$ We propose the first work that multiplexes several tasks on a single physical representation. \FW  orthogonalizes the task representations at the semantic level, thus enabling their joint processing and transmission to the receiver which demultiplexes the tasks and complete the processing to obtain the results. This way, \FW drastically reduces the overall end-to-end latency due to task execution; \smallskip

\noindent $\bullet$ We present a novel semantic channel sounding strategy for dynamically adapting the \FW transmitter and receiver in the inference phase to preserve semantic orthogonality among computation and communication channels, accounting for the varying wireless environment; \smallskip

\noindent $\bullet$ We extensively evaluate \FW through real-world experiments. We prototyped \FW using \gls{mimo} \glspl{sdr} as the transmitter and receiver radio frontends and a Jetson Orin Nano board as the mobile device. We evaluated \FW experimentally and compare it with state-of-the-art approaches in \gls{semcom} on two tasks (image classification and sentiment analysis) to evaluate its efficacy on different input types (i.e., images vs text) and using different processing functions (Residual networks vs Transformers). Our experiments on system scalability show that \FW allows multiplexing 8 tasks on 4 communication channels with a drop in accuracy of less than 2.5\% with respect to multiplexing 4 tasks. \FW decreases the end-to-end latency (communication plus computation) with respect to state-of-the-art semantic strategies by more than 7$\times$ on average and reduces energy consumption and communication load by more than 25$\times$ and 54$\times$ respectively.\smallskip

\noindent \textbf{Summary of Impact}\vspace{0.1cm}

\noindent For the first time, this paper introduces and validates a new concept: \textit{the number of tasks that can be semantically multiplexed in a communication system is greater the number of available physical streams}. As such, we believe  this paper opens new and exciting research avenues at the intersection of machine learning and wireless networking. \vspace{-0.2cm}

\section{Semantic Multiplexing}\label{sec:framework}

As depicted in Figure~\ref{fig:overview}, \FW semantically multiplexes multiple inputs for efficient transmission over wireless channels. Since the semantic transmission and reception pipelines are jointly optimized,  \FW is independent from the specific number of antennas at the transmitter and receiver. Specifically, the \FW transmitter maps the task inputs into latent representations to be directly transmitted through the available antennas, while the \FW receiver processes the received signals to obtain the task outcomes.\vspace{-0.2cm}

\begin{figure}[!h]
    \centering
    \includegraphics[width=\linewidth]{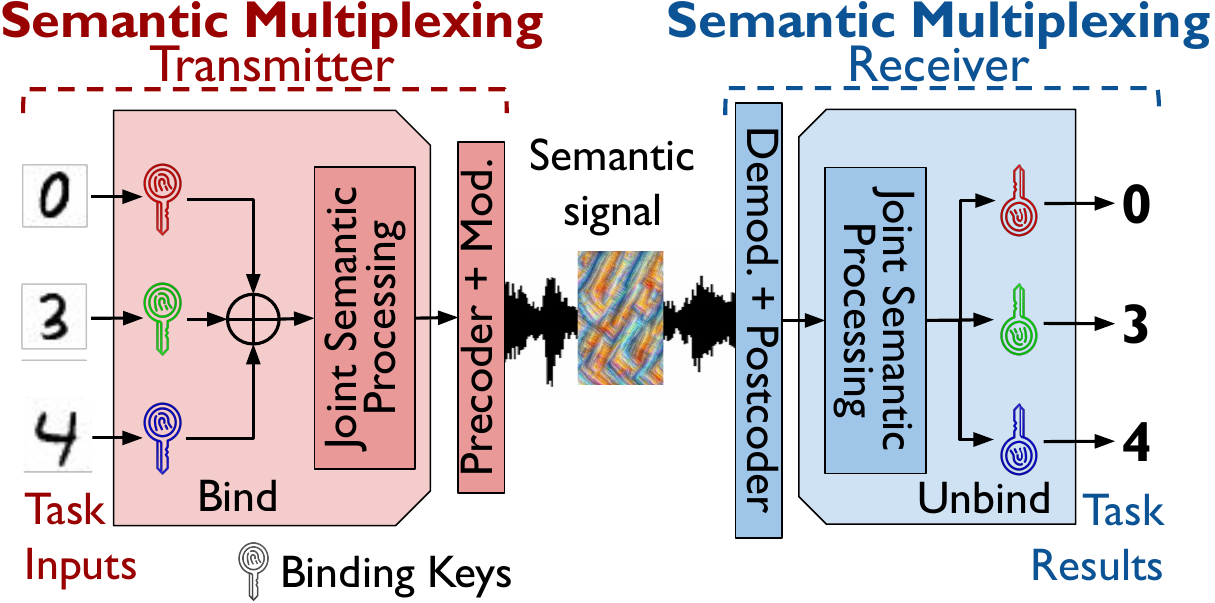}
    \setlength\abovecaptionskip{-0.2cm}
    \caption{\FW transmitter and receiver. `Mod.' and `Demod.' stand for modulator and demodulator.\vspace{-0.25cm}}
    \label{fig:design}
\end{figure}

The \FW transmitter is summarized on the left side of Figure~\ref{fig:design}. The first operation consists of \emph{binding} the task inputs using specific protected keys (\textit{see Section \ref{sec:binding}}). This operation ensures that each input is processed in a separate protected subspace thus enabling semantic multiplexing~\cite{murahari2023mux, murahari2022datamux, menet2023mimonets}. Next, the bound key-value pairs are summed up element-wise to form a single compositional data structure that undergoes joint semantic processing (\textit{see Section \ref{subsec:joint_processing}}). The output is then fed to learnable task-oriented precoding module, which allows \FW to react quickly to sudden changes in the wireless channel. Based on \gls{csi}, this module precodes the latent symbols by learning to optimize the end task performance (\textit{see Section \ref{subsec:stocastic-precoder}}). Finally, a non-trainable modulator generates the output waveform (\textit{see Section \ref{subsection:modulator}}). At the \FW receiver (see the right side of Figure~\ref{fig:design}), a non-trainable demodulator is combined with a data-driven task-oriented postcoding module and a joint semantic  processing block to directly process the incoming   waveforms to execute the task. The superposed task outputs are demultiplexed performing the unbinding operations using the corresponding keys. 

As detailed in Section~\ref{sec:math}, the \FW transmitter and receiver are jointly trained by adding a non-trainable stochastic model that represents the communication channel to link the transmitted and received latent representations. The rest of this section details the \FW transmitter and receiver modules.\vspace{-0.25cm}

\subsection{Binding/Unbinding}\label{sec:binding}

The binding mechanism used by \FW is based on \gls{hrr}, which are implemented using circular convolutions~\cite{plate1995holographic}. More precisely, circular convolution is repeatedly applied between a binding key and each input volume spanning across the feature maps. As a result, binding is translation-equivariant and maintains locality. The unbinding mechanism, on the other hand, is based on \gls{mbat} and is implemented through matrix multiplication~\cite{gallant2013representing}. The protection keys for binding and unbinding are initialized randomly and leaned during training to mitigate the interference between communication channels. For this, we leverage the fact that random tensors become quasi-orthogonal as their dimension increases. Therefore, we use unique high-dimensional keys to form quasi-orthogonal key-value pairs that can coexist and be processed concurrently. The binding keys are data-independent and are only linked to the specific multiplexed task input (i.e., computation channel). We refer interested readers to \cite{kleyko2022survey} for alternative binding mechanisms.\vspace{-0.25cm}

\subsection{Joint Processing}\label{subsec:joint_processing}

After the binding operation, the task inputs undergo joint semantic processing to reduce latency and energy consumption. We model this operation as a differentiable function $f_t(\omega_t)$ that processes the bound tasks toward the task execution. At the receiver, the joint processing, defined as a differentiable function $f_r(\omega_r)$, completes the task execution before the unbinding. Note that if the functions are not differentiable, they should be re-parametrized as differentiable functions, e.g., using~\cite{jermyn2005invariant,jang2022reparametrization}, to be included in the joint transmitter and receiver semantic optimization through loss backpropagation as detailed in Section~\ref{sec:math}.

Both learnable and heuristic functions can be used within \FW depending on the task to be executed. If learnable, these modules help the precoding and postcoding in generating latent representations that are robust to the channel modifications and compensating for additional channel impairments at the receiver, as discussed in Section~\ref{subsec:stocastic-precoder}.

To help task execution, two additional disjoint processing steps can be integrated into the system. At the transmitter, before binding, the task inputs may be pre-processed to obtain higher-dimensional representations which further help isolate their processing once bound together. At the receiver, the unbound outputs may undergo another disjoint processing step to refine the task results. Whether to include these additional steps or not is a design choice and depends on the specific task to be executed and the computational power of the transmitter and receiver. As shown in Section~\ref{subsec:scalability_evaluation}, there is a tradeoff between the performance gain and the overall complexity of the system to account for in the system design.\vspace{-0.1cm}

\begin{figure}[h]
    \centering
    \includegraphics[width=\linewidth]{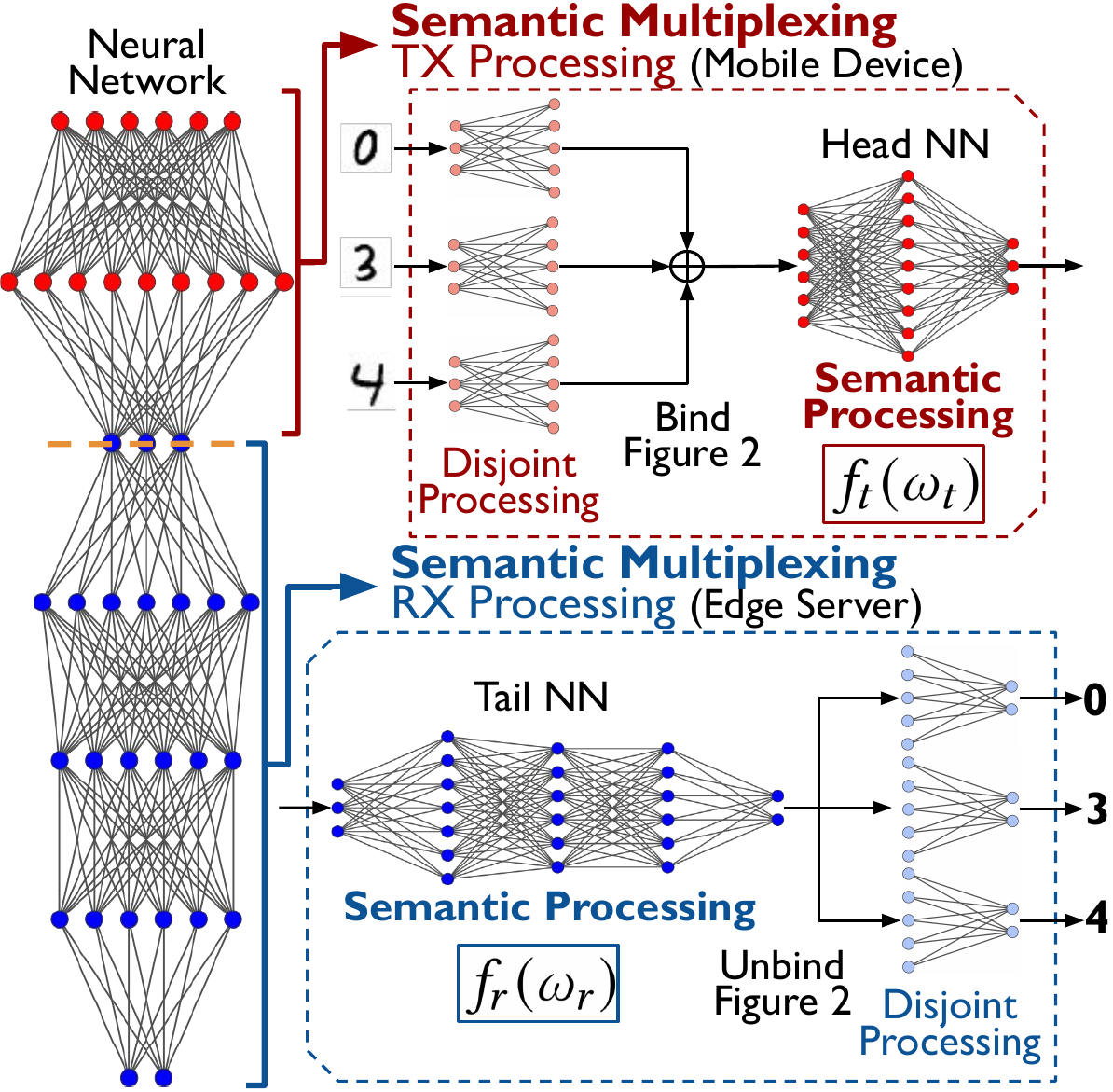}
    \setlength\abovecaptionskip{-0.2cm}
    \caption{Example of \FW processing using a DNN for task execution.\vspace{-0.25cm}}
    \label{fig:processing}
\end{figure}

\smallskip
\noindent \textbf{Example of Joint Processing Functions.}~Our modeling applies to several tasks based on neural networks, including object detection with Residual networks~\cite{zagoruyko2016wide} and text analysis with Transformer networks~\cite{maas-etal-2011-learning}. As shown in Figure~\ref{fig:processing}, the \gls{dnn} responsible for task execution can be divided into two differentiable and learnable processing functions, which are then deployed at the \FW transmitter ($f_t(\omega_t)$) and receiver ($f_r(\omega_r)$) \cite{kang2017neurosurgeon,mohammed2020distributed, liang2023dnn}. At the transmitter, a portion of the head \gls{dnn}, e.g., the first layer, can also be replicated for all inputs and used as the disjoint preprocessing function that helps the binding mechanism (see Figure~\ref{fig:processing}, top). At the receiver, a similar strategy can be applied by isolating and replicating the last layer(s) for a final processing step on the unbound outputs to obtain the task results (see Figure~\ref{fig:processing}, bottom). For example, in a classification task, the last linear (classifier) layer can be used.

\subsection{Precoding/Postcoding}\label{subsec:stocastic-precoder}

The precoding and postcoding allow quickly responding to channel variations maintaining sufficient task performance.  \vspace{0.1cm}

\noindent \textbf{Design Choices.}~We design a learnable \textit{stochastic precoder} since deterministic strategies result in a latent space where closer points do not necessarily associate with closer points in the task output space. As such, since deterministic models memorize an exact mapping between the latent and the output, a small perturbation of the latent -- whether intentional (adversarial) or unintentional (channel distortions) -- can change the task result drastically. Conversely, using a stochastic precoder with a customized loss function (see Section~\ref{subsec:loss}) allows imposing a prior distribution on the latent symbols, thus increasing the generalization ability. A deterministic module is used for the learnable postcoder to simplify the formulation. 

We examined three different approaches to design the precoder (the same applies ``mirrored'' to the postcoder). The most straightforward approach entails conditioning the entire transmitter processing function on the \gls{csi}, i.e., defining $f_t(\omega_t | s)$ to process the bound task inputs. However, this increases the processing function complexity (e.g., the input layer size if a \gls{dnn} is used as the processing function), incurring a high computation overhead on the transmitter, which is likely to be an energy-constrained mobile device as detailed in Section~\ref{sec:intro}. Another option is to add another processing function (e.g., a separate neural network) to the transmitter to jointly process the output of the bound input processing (through function $f_t(\omega_t)$) and the \gls{csi}. Although this option leaves the original transmitter function $f_t(\omega_t)$ intact, it still leads to an increase in the transmitter computation load. Hence, we decided to use use a dedicated precoding module to process the \gls{csi} separately, compute a task-oriented precoding tensor, and use it to transform the processed latent symbols to compensate for channel distortion. This option allows trading off representational power with reduced computation load.\vspace{0.1cm}

\begin{figure}[h]
    \centering
    \includegraphics[width=0.92\linewidth]{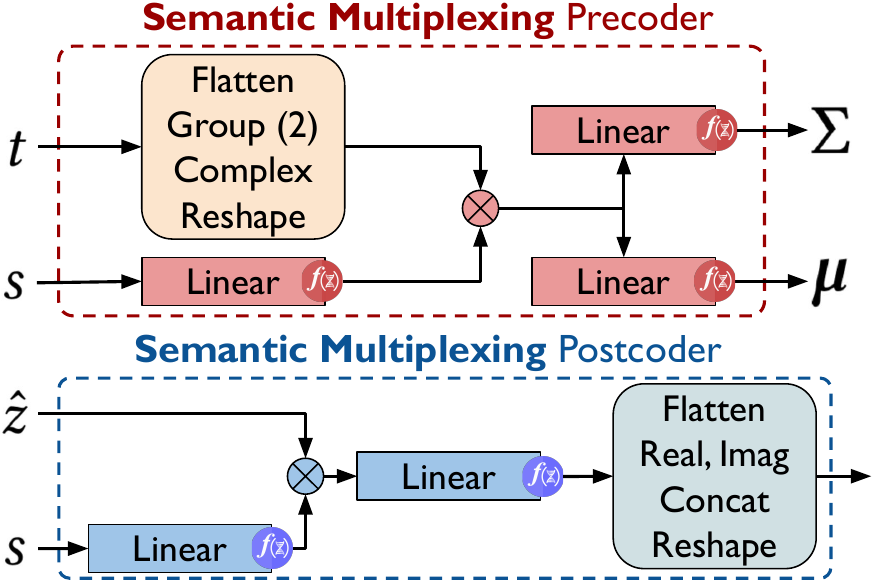}
    \caption{Semantic precoding/postcoding in \FW.\vspace{-0.3cm}}
    \label{fig:precode-design}
\end{figure}

\noindent \textbf{Design Details.}~The precoder structure is depicted in the top side of Figure~\ref{fig:precode-design}. As further detailed in Section~\ref{sec:math}, we assume that the conditional probability of latent symbols given the input has a complex Gaussian distribution, and the transmitter processing at the transmitter $f_t(\omega_t)$ combined with the precoder extract the mean and covariance of this distribution. Let $t$ be the bound and processed inputs obtained as output from $f_t(\omega_t)$. Since the \FW precoder processes complex \gls{i/q} symbols, tensor $t$ is first converted to a complex tensor using the first part as the real and the second as the imaginary parts. The resulting tensor is split into multiple packets. The \gls{csi}, denoted by $s$, is fed to a linear layer with non-linear activation to compute the semantic precoding tensor for each packet. After applying the precoding (matrix multiplication) the packets are fed to other two linear heads with non-linear activations to extract the mean ($\boldsymbol{\mu}$) and the covariance ($\boldsymbol{\Sigma}$). The latent symbols generated by sampling a Gaussian distribution with these parameters and passed to the modulator described in Section~\ref{subsection:modulator} to generate the waveforms ready to be transmitted.

The postcoder (Figure~\ref{fig:precode-design}, bottom) calculates the semantic postcoding tensor for each antenna similarly to the precoder obtaining the semantic postcoder tensor by processing the \gls{csi} $s$ through a learnable linear transformation with a non-linear activation function. The corrupted latent symbols are postcoded using this tensor, processed through another linear layer with a non-linear activation, converted to real-valued feature maps and fed to the receiver processing block consisting of joint processing $f_r(\omega_r)$, unbinding and optional disjoint processing to obtain the tasks outcomes.

\subsection{Modulation/Demodulation}\label{subsection:modulator}

At the transmitter, a modulator takes the latent symbols (from the precoder) in the frequency domain and transforms them to the time-domain baseband symbols, which can be up-converted to the carrier frequency and transmitted via a radio frontend. On the other hand, the received waveforms are down-converted to \gls{i/q} samples and further transformed to frequency-domain symbols by a demodulator. As such, the latent symbols transmission can be modeled as a matrix multiplication between the latent symbols at the modulator input (frequency domain) and the \gls{cfr}, followed by noise that provides the distorted latent symbols at the demodulator output. In this paper, in line with modern communication systems, we have realized the modulator/demodulator with direct and inverse \gls{fft} operations.\vspace{-0.2cm}

\section{Mathematical Optimization}\label{sec:math}

The \FW system described in Section~\ref{sec:framework} is jointly trained to enable task-oriented precoding and postcoding that jointly optimize the communication and computation objectives. This is one of the key challenges in \gls{semcom}~\cite{cai2024multi}. For this reason, previous work separately optimizes communication and computation thus obtaining a sub-optimal system without semantic multiplexing capabilities (e.g., see the MCR2/sm baseline~\cite{cai2024multi,cai2025end} in Section~\ref{sec:exp-result}). In this section, we detail the new information-theoretical framework we developed to enable the combined training of all the learnable blocks in \FW. One of the key innovation we introduce is the integration into the system of a channel model which describes how the signals are modified during wireless propagation (see Section~\ref{subsec:channel-model}). Once derived the complete probabilistic model, we use the loss function formulated in Section~\ref{subsec:loss} to jointly train the \FW modules. In Section~\ref{subsec:adapt} we detail the strategies we designed and implemented to adapt dynamically the \FW system to compensate for the changing wireless channel. 

\smallskip
\noindent\textbf{Notation.} Random variables and their realizations are indicated by upper-case and lower-case letters, respectively. $E\{X\}$ and $H(X)$ denote the statistical expectation and entropy of variable $X$. The mutual information between $X$ and $Z$ is indicated by $I(X; Z)$, while $H(Z| X)$ is the conditional entropy of $Z$ given $X$. We use $D_{KL}(p(x) || q(x))$ for the \gls{kl} divergence between $p(x)$ and $q(x)$ probability distributions. Finally, $\mathcal{CN}\!(\!\boldsymbol{\mu}, \mathbf{\Sigma})$ refers to the circularly-symmetric complex Gaussian distribution with mean $\boldsymbol{\mu}$ and covariance matrix $\mathbf{\Sigma}$.

\begin{figure}[t]
  \centering
  \includegraphics[width=\linewidth]{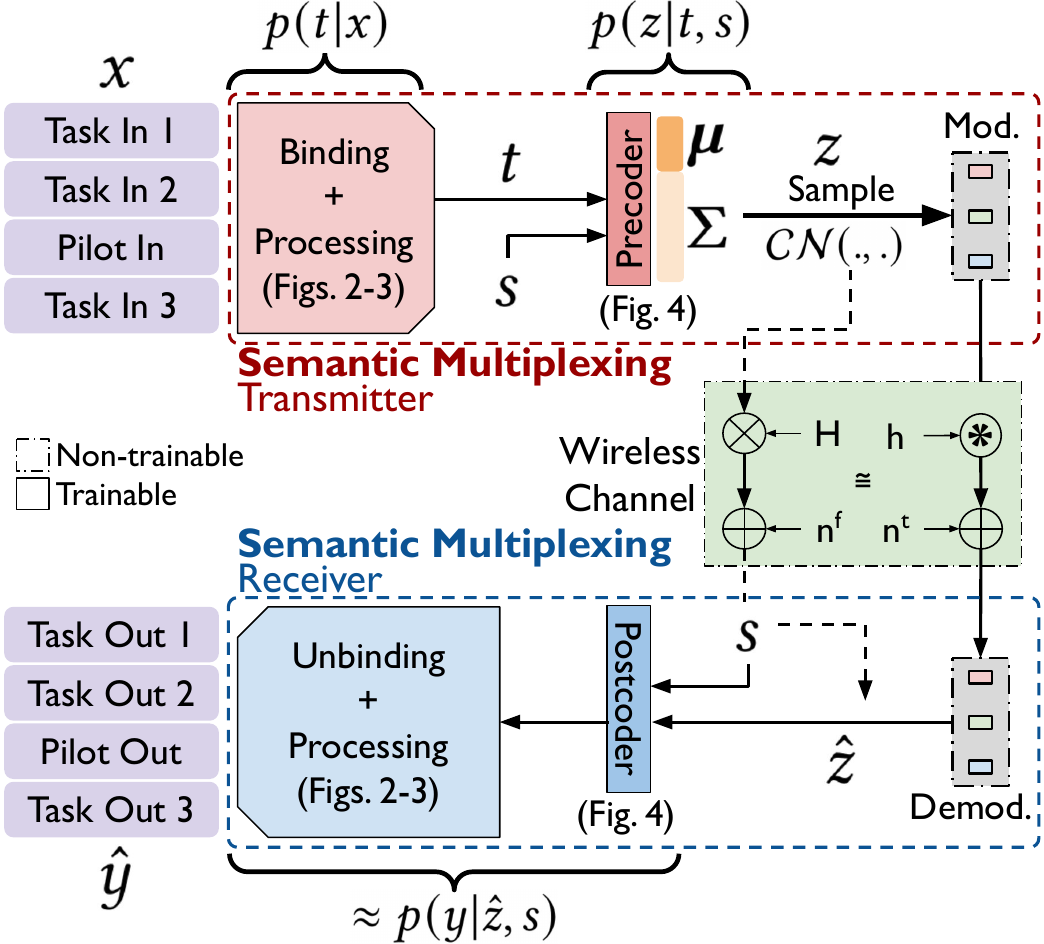}
    \setlength\abovecaptionskip{-0.1cm}
  \caption{\FW probabilistic system model.\vspace{-0.4cm}}
  \label{fig:sys_model}
\end{figure}

\smallskip
\noindent\textbf{Model Walkthrough.} The probabilistic model for \FW is illustrated in Figure~\ref{fig:sys_model}. We assume that the task input data $x$ and their corresponding desired outputs $y$ are generated by an information source having an underlying joint probability distribution $p(x, y)$. Let $z$ be the latent symbols which are the frequency-domain input to the modulator. In the information-theoretic sense, we want these symbols to contain only the information needed to complete the task and obtain $y$ at the receiver. The extraction of such symbols $z$ from the inputs $x$ and conditioned on the \gls{csi} ($s$) for each packet $p$ and subcarrier $k$, is modeled as sampling from the conditional probability $p(z_{p, k}| x, s)$. This probability is assumed to have a Gaussian distribution, which parameters are found combining the transmitter processing together with the stochastic precoder described in Section~\ref{subsec:stocastic-precoder}. To reduce the computational load, we adopt the mean-field assumption that the conditional probabilities for different packets and subcarriers in $z$ are independent~\cite{blei2017variational}, i.e., $p(z| x, s) \!=\! \prod_{p = 1}^{P} \prod_{k = 1}^{K} p(z_{p, k}| x, s)$. This is a common assumption in multi-carrier \gls{mimo} systems, where each packet and subcarrier is processed independently. Let $f_{p, k}^{\boldsymbol{\mu}}(x, s; \phi)$ and $f_{p, k}^{\boldsymbol{\Sigma}}(x, s; \phi)$ be the functions implemented by the combination of the transmitter processing ($f_t(\omega_t)$) and the precoder for obtaining $\boldsymbol{\mu}$ and $\boldsymbol{\Sigma}$, respectively, for packet $p$ and subcarrier $k$, with $\phi$ denoting the learnable parameters, $p(z| x, s)$ writes as\vspace{-0.15cm}
\begin{equation}
\begin{split}
    \label{eqn:latent-cdist}
    p(z| x, s) = \prod_{p = 1}^{P} \prod_{k = 1}^{K} \mathcal{CN}\bigl(f_{p, k}^{\boldsymbol{\mu}}(x, s; \phi), f_{p, k}^{\mathbf{\Sigma}}(x, s; \phi)\bigr).
\end{split}\vspace{-0.15cm}
\end{equation}

The conditional probability $p(z_{p, k}| x, s)$ is approximated following two phases. At first, the transmitter precoding obtains $p(t| x)$ conditioned on the inputs, where $t$ is the processing output representation. Second, the stochastic precoder implements $p(z_{p, k}| t, s)$ conditioned on $t$ and the \gls{csi}, obtaining $p(z_{p, k}| x, s)\! =\! \int p(z_{p, k}| t, s) p(t| x) dt$.
After deployment, the transmitter modulates the symbols sampled from $p(z| x, s)$ and transmits the waveforms over the channel. \smallskip

The receiver collects the distorted signals and demodulates them into corrupted latent symbols $\hat{z}$ which are then fed to the postcoder plus the receiver processing block ($f_r(\omega_r)$) to finalize task inference and compute the results $\hat{y}$, using the function $g(\hat{z}, s; \theta)$, where $\theta$ denotes the learnable parameters. The prediction of task outputs $y$ from the corrupted latent symbols $\hat{z}$ and conditioned on \gls{csi} $s$ is modeled as sampling from a conditional probability $p(y| \hat{z}, s)$. \vspace{-0.2cm}

\subsection{Modeling the Wireless Channel}\label{subsec:channel-model}

During training, \FW uses a stochastic model for signal propagation, representing the wireless channel between the transmitter and receiver in the frequency domain. The model accounts for the different number of antennas (one/multiple) at the transmitter and the receiver. We consider frequency-selective multi-path fading effects plus an additive complex white Gaussian noise. Both Rayleigh and Rician distributions are investigated for the fading. The \gls{csi} $s$ modeling the channel includes both the \gls{cfr} matrix $H$ and the noise variance vector in the frequency domain denoted by $\sigma^2$. As shown in Figure~\ref{fig:design}, the latent symbols obtained at the transmitter are multiplied by the \gls{cfr} $H$. Noise $n^f \!\sim\! \mathcal{CN}(0, \sigma^2)$ is then added to obtain the corrupted latent symbols. Considering the constraints on the transmitter power, we normalize the peak power of the modulated signal to $1$, resulting in an \gls{snr} of $10 \log_{10} \frac{1}{\sigma^2}$. The modulator, demodulator, and channel model are non-trainable yet differentiable layers during the end-to-end training of \FW, and are modeled by the equivalent conditional distribution $p(\hat{z}| z, s)$.

Accounting for the wireless transmission in \FW is needed because the latent representations are transmitted as \gls{phy} waveforms. Hence, each latent collected at the receiver is a channel-distorted version of the transmitted representation. During the end-to-end training described in Section~\ref{subsec:loss}, \FW identifies the distortion patterns caused by the channel and learns a custom channel coding strategy that robustifies the latent features against them. As such, \FW performs channel coding on top of its processing task.\vspace{-0.2cm}

\subsection{Formulation of Loss Function} \label{subsec:loss}

As introduced before, from an information-theoretic viewpoint, our goal is to learn a latent $Z$ that captures only the task-relevant information in $X$ to predict $Y$ at the receiver. Hence, we formulate the loss function used to train \FW following the \gls{ib} principle~\cite{tishby2000information} as\vspace{-0.05cm}
\begin{equation}
\begin{split}
    \mathcal{L}_{IB} =&-I(\hat{Z}; Y| s) + \beta \cdot I(X; Z| s)\\
    = &~ \mathbb{E}_{p(x, y)} \biggl\{\mathbb{E}_{p(s)} \Bigl\{\mathbb{E}_{p(\hat{z}| x, s)} \bigl[-\log p(y| \hat{z}, s)\bigr] \\
    &+\beta \cdot D_{KL}\bigl(p(\hat{z}| x, s) || p(z)\bigr)\Bigr\}\biggr\} - H(Y),
    \label{eqn:loss_ib}
\end{split}\vspace{-0.2cm}
\end{equation}
where $0 \!\leq \!\beta \!\leq \!1$ is the weight controlling the efficiency-performance trade-off.
By minimizing this loss function, we force $Z$ to forget the unnecessary information in $X$ by minimizing the mutual information $I(X; Z| s)$, and retain only the minimal sufficient statistics of $X$ for expressing $Y$ by maximizing the mutual information $I(\hat{Z}; Y| s)$. As a result, we can find the optimal $Z$ which is maximally compressive about $X$ while being maximally predictive about $Y$.

The conditional distribution $p(\hat{z}| x, s)$ in Equation~\eqref{eqn:loss_ib} is obtained as $p(\hat{z}| x, s) \!= \!\int p(z| x, s) p(\hat{z}| z, s) dz$, where $p(z| x, s)$ is derived in Equation~\eqref{eqn:latent-cdist}. However, computing the remaining distributions $p(z)$ and $p(y| \hat{z}, s)$ for high-dimensional data with arbitrary distributions is computationally prohibitive~\cite{abdi2025phydnns}. 
To overcome this issue, we leverage the \gls{vib} theory~\cite{alemi2016deep} to obtain a variational upper bound of the loss and the Monte Carlo sampling to derive an unbiased estimation of the expected values in the expression. Note that the precoder output is a distribution, and it should be sampled to obtain a realization. However, such sampling is not differentiable. Therefore, we use the reparameterization trick \cite{kingma2013auto} to make \FW trainable through loss backpropagation using a gradient descent algorithm. The details of this loss reformulation are included in Appendix~\ref{subsec:variational_bound}.\vspace{-0.2cm}

\subsection{Dynamic Adaptation}\label{subsec:adapt}

\begin{figure}[h]
    \centering
    \includegraphics[width=\linewidth]{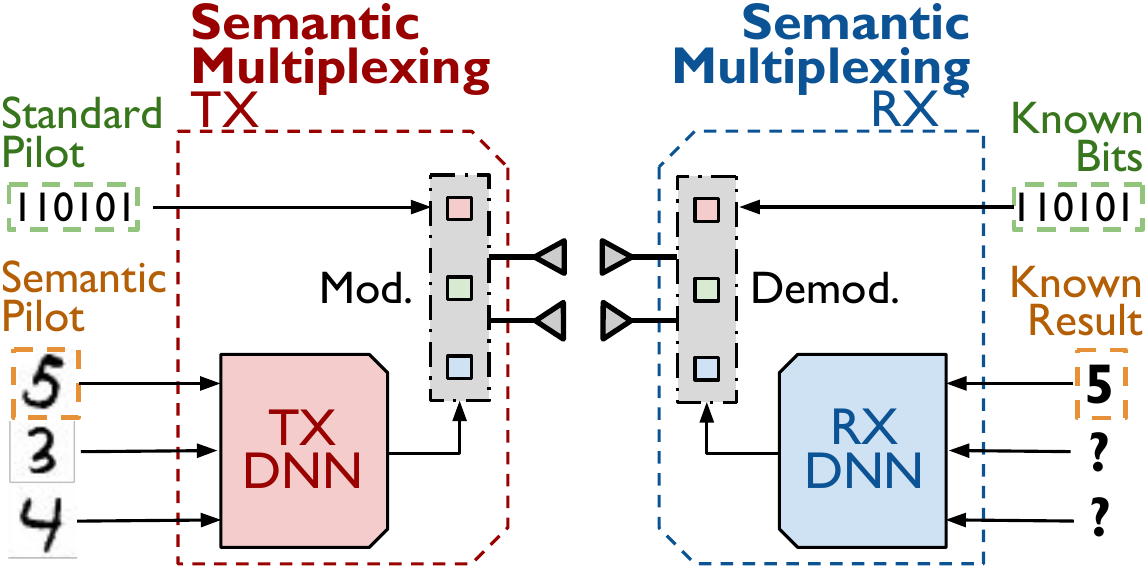}
    \setlength\abovecaptionskip{-0.05cm}
    \caption{Standard and semantic channel sounding.\vspace{-0.5cm}}
    \label{fig:pilots}
\end{figure}

After deployment and during testing, \FW alternates between two operation modes to sound the wireless channel and adapt to changing conditions as summarized in Figure~\ref{fig:pilots} and described next. While the first sounding is inherited from standard communication procedures, the second adaptation strategy is specifically designed for \FW framework and has never been proposed before.\vspace{-0.2cm}

\smallskip
\noindent \textbf{Communication Pilots Transmission.}~Each time before executing the inference, \FW triggers the standard channel sounding procedure adopted for precoding in \gls{mimo} networks~\cite{goldsmith_2005}. As such, the mobile device transmits a packet containing pilot symbols only. Using them, the edge server estimates the \gls{csi} $s$ and feeds it back to the mobile device. The estimated \gls{csi} is used by the precoding and postcoding to make \FW adapt to the varying wireless channel.

\smallskip
\noindent \textbf{Task-Oriented Pilots Processing.}~During normal operation mode, \FW randomly selects a computation channel to process task-oriented pilots. As the transmitter and receiver have to stay synchronized in selecting the index of such a computation channel, \FW uses a hash function to obtain the index depending on the global time and a shared seed (established beforehand) without the need for explicit communication. The transmitter uses the selected channel to process an input with a known task output. For this, the transmitter device has to store a tiny subset of the training inputs as task-oriented pilots. Since the task result of the selected computation channel is known by the edge server, the corresponding loss can be calculated and backpropagated through that channel. Therefore, the \FW transmitter and receiver \glspl{dnn} are adapted in an end-to-end manner taking into account the updated \gls{mimo} channel model using the \gls{csi} estimated through communication pilots. This test-time adaptation procedure maintains isolation between different computation and communication channels.\vspace{-0.1cm}

\section{Experimental Prototype}\label{sec:exp-eval}

\begin{figure}[t]
    \centering
    \includegraphics[width=\linewidth]{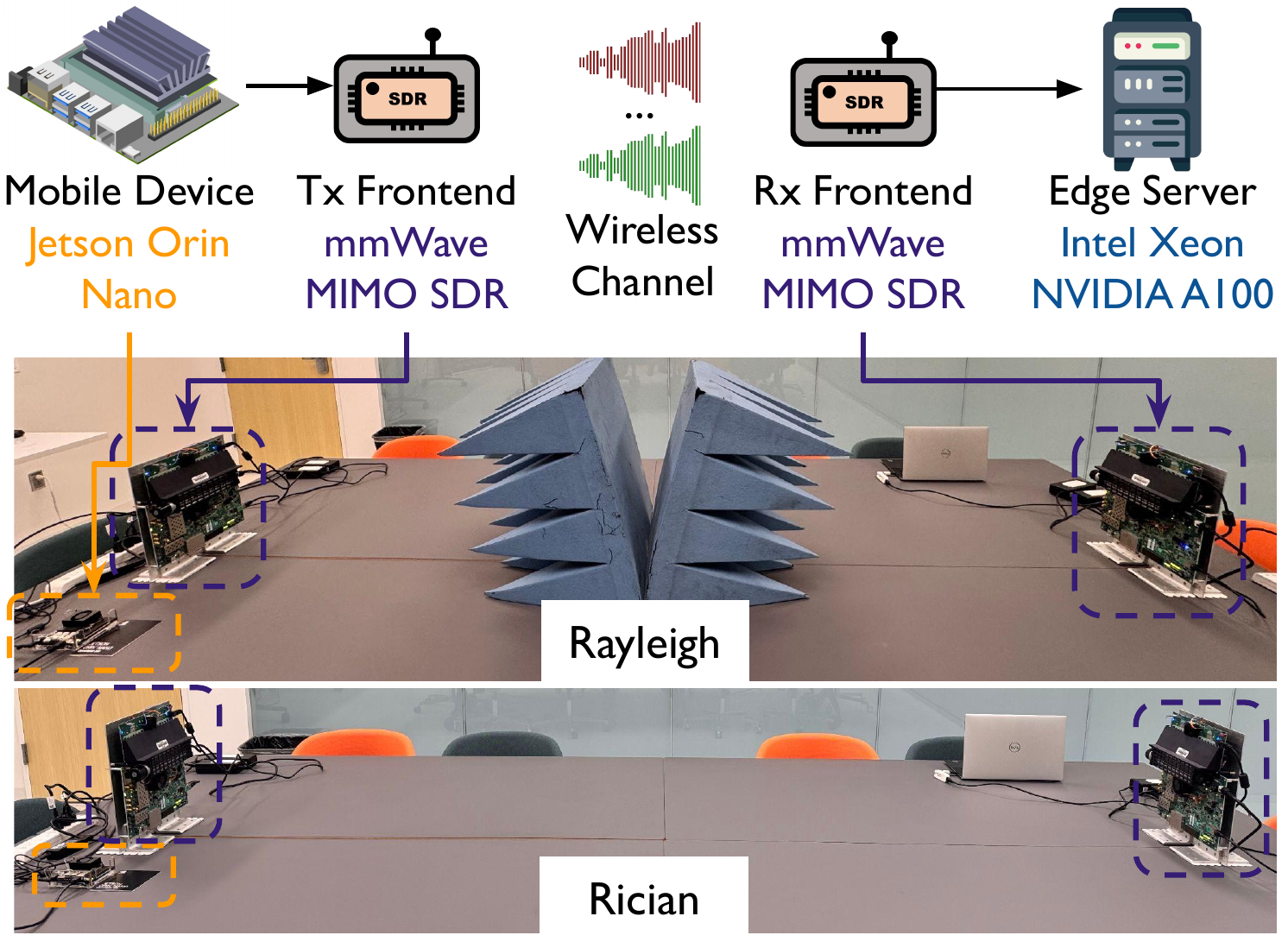}
    \setlength\abovecaptionskip{-0.2cm}
    \caption{Experimental setup for real-world evaluation of \FW in a conference room in NLoS (Rayleigh fading, with the absorbers) and LoS (Rician fading).\vspace{-0.55cm}}
    \label{fig:exp-setup}
\end{figure}

We developed an end-to-end prototype illustrated in the upper part of Figure~\ref{fig:exp-setup}, where we use a Jetson Orin Nano board powered by a 32-tensor-core NVIDIA Ampere GPU and a 6-core ARM Cortex-A78AE CPU as the (energy-constrained) transmitter. A workstation equipped with a 432-tensor-core NVIDIA A100 GPU and a 12-core Intel Xeon Silver 4410Y CPU is considered as the  receiver. \Gls{mmwave} \glspl{sdr} are used as the transmitter and receiver. The \glspl{sdr} operate at 60 GHz with a bandwidth of 1~GHz and are equipped with 8 digitally-controlled RF chains each, thus supporting fully digital 8 $\times$ 8 \gls{mmwave} communications. The Jetson Orin Nano board is connected to the transmitter radio while the other \gls{sdr} is connected to the receiver. The prototype has been deployed  in a conference room, as shown at the bottom of Figure~\ref{fig:exp-setup}. We conducted the experiments both with and without the absorber foams placed at 1.5~m distance from the radios, corresponding to \gls{nlos} and \gls{los} scenarios (Rayleigh and Rician fading). \vspace{0.1cm}

During training, we consider 20~dB \gls{snr} to determine the additive noise variance in the probabilistic model while the \gls{cfr} is collected from the experimental setup and changed at every epoch. We incorporated an \gls{ofdm} modulation scheme with an \gls{fft} size of 1024, where 800 subcarriers are used out of 1024, while the rest are left unused as guard bands. ReLu is used as the non-linear activation function in the precoder and postcoder. The trade-off parameter in Equation~\ref{eqn:loss_mc} is set to $\beta\!=\!10^{-4}$. In the testing phase and during adaptation, all the parameters are frozen, except for the precoding and postcoding parameters plus the binding and unbinding keys.\vspace{-0.1cm}

\subsection{Processing Tasks}\label{subsec:network}

We tested \FW with image classification and sentiment analysis tasks to evaluate the efficacy of \FW on different input types (i.e., images vs text) and using different processing functions (Residual networks vs Transformers). \vspace{0.1cm}

\noindent\textbf{Image classification.}~Consistent with existing work, we considered CIFAR-10, CIFAR-100, and SVHN. CIFAR-10 and CIFAR-100 contain 50,000 training images and 10,000 testing images each, distributed evenly among 10 and 100 classes, respectively. SVHN comprises 73,257 training images and 26,032 testing images across 10 classes. All datasets have RGB input channels. Data augmentation was consistently applied to the training set through random horizontal flips, random cropping, and normalization. WideResNet \glspl{dnn} were selected for image classification. Specifically, we selected the learning models WRN-28-10, WRN-16-8, and WRN-10-4~\cite{zagoruyko2016wide}, where the first and second numbers in the name indicate the depth (number of layers) and width (number of kernels) of each model. The larger architectures provide improved performance at the cost of increasing computational and memory load. We divide each backbone after the first ResNet block and deployed the one-block head \gls{dnn} on the mobile device (joint processing function $f_t(\omega_t)$ in Section~\ref{subsec:joint_processing}) and the remaining tail \gls{dnn} at the edge server ($f_r(\omega_r)$), as depicted in Figure~\ref{fig:processing}. \vspace{0.1cm}

\noindent\textbf{Sentiment analysis.} We considered the second task in the \gls{lra} Transformers benchmark on text analysis~\cite{tay2021long}. We used the IMDb reviews dataset, which contains 50,000 movie reviews for binary sentiment classification~\cite{maas-etal-2011-learning}. The authors of the dataset defined a set of 25,000 highly polar movie reviews for training and 25,000 for testing. We used the Transformer architecture in~\cite{menet2023mimonets} as the signal processing function for sentiment classification. Specifically, the selected architecture consists of 6 attention blocks with 8 heads each. This architecture presents an additional challenge with respect to the WideRedNet as the binding and unbinding processes should be sequentially performed for each of the of the attention blocks to enable task multiplexing. We deployed the first attention layer until the multi-layer perceptron before the unbinding at the transmitter, while the unbinding for the first block and the remaining attention blocks have been deployed at the receiver.

\vspace{0.1cm}
For both tasks, at the transmitter, the first convolutional layer is cloned for each of the task inputs as a preliminary disjoint processing to improve the separation in the semantic space enforced by the binding mechanism. At the receiver, the last linear (classifier) layer is duplicated for all the outputs. Each classifier is fed with one of the unbound values and returns the associated task result.
\vspace{-0.2cm}

\subsection{Baseline Approaches}\label{subsec:benchmarks}

We consider four baselines. For baseline naming we use the convention `MethodName/xy', where `x' and `y' identify the computation and communication strategy, respectively, adopted by the method, which can be multiplexed (letter m) or not (letter s). Notably, \FW is the first approach in the literature proposing to multiplex both computation and communication. In the following, we describe the baselines by grouping them into two categories. The first includes approaches that use semantic at the application layer and transmit data using full protocol stack, while the approaches in the second directly obtain \gls{phy}representations. \vspace{0.1cm}

\noindent \textbf{Full-Stack \gls{semcom}:}~These approaches use IEEE 802.11 (Wi-Fi) to transfer latent representations. Specifically, the latent floating-point features for each batch of images are converted into their bit representations and divided into application layer \glspl{pdu} based on a suitable \gls{mtu} size. The \glspl{pdu} are encapsulated using \gls{tcp} and the Wi-Fi data link layer to form physical-layer bits.
The \gls{i/q} symbols are obtained using a \gls{qpsk} modulation scheme and fed to an \gls{ofdm} block to be further transmitted using the radio. The retransmission mechanisms employed at different layers of the protocol stack allow these methods to maintain performance with different communication channel conditions, at the cost of increasing end-to-end latency and mobile energy consumption. We consider the following baselines:\vspace{0.1cm}

$\bullet$ \textbf{MNet/ms:}~We used the method proposed in~\cite{menet2023mimonets} to build \gls{mimo}-\glspl{dnn} based on WideResNets. Then, we applied the algorithm in \cite{kang2017neurosurgeon} to divide the \glspl{dnn} between the transmitter and the receiver. The first layer of the \gls{dnn} is deployed on the transmitter, while the rest is executed on the receiver.\vspace{0.1cm}

$\bullet$ \textbf{BF/ss:}~The authors in~\cite{matsubara2022bottlefit} reduced the amount of transferred data learning a compressed latent representation in single-input \glspl{dnn} by reducing the width of the last convolutional layer in the first WideResNet block and other affected layers. A two-stage training utilizing \gls{kd} is used to compensate for accuracy drop. We followed the same approach and used 9 and 3 kernels as the reduced width, indicated as BF-9/ss and BF-3/ss, respectively.

\vspace{0.2cm}
\noindent \textbf{Physical Layer \gls{semcom}:}~In these approaches, the latent symbols are already channel-encoded. We implemented the packet detection and synchronization similarly to Wi-Fi. These methods implicitly compensate for channel impairments regardless of the wireless condition and without relying on retransmissions. We consider two baseline methods:\vspace{0.1cm}

$\bullet$ \textbf{MCR2/sm:}~In \cite{cai2024multi,cai2025end}, a \gls{mimo} \gls{semcom} system uses linear precoding and postcoding blocks similar to those in traditional communication systems, which are updated through an iterative algorithm. Class-wise separability is adopted as a surrogate measure for classification accuracy. The entire \gls{dnn} up to the penultimate layer is executed at the transmitter which sends only the class probabilities to the receiver. While in the provided implementation a fixed \gls{csi} is used, we made it adaptive to have a fair comparison with \FW.\vspace{0.1cm}

$\bullet$ \textbf{PhyDNN/ss:}~\Citet{abdi2025phydnns} proposed modifying and fine-tuning an already-trained \gls{dnn} to deploy it directly in the \gls{phy} layer producing single-stream transmissions. Their key modification involves introducing a codebook into the \gls{dnn} to produce discrete latent symbols compatible with a given digital modulation scheme. To utilize this approach as a baseline, we applied it to WideResNet models, generating \gls{ofdm} symbols at the output of the first layer.\vspace{-0.2cm}

\section{Experimental Results}\label{sec:exp-result}

In Section~\ref{subsec:task_accuracy}, we study the performance of \FW in terms of task accuracy. We show that the joint design of computation and computation does not lead to consistent performance degradation with respect to full-stack approaches and allow increasing accuracy for \gls{phy} strategies. In Section~\ref{subsec:scalability_evaluation}, we study the scalability of \FW when multiplexing an increasing number of task inputs (i.e., computation channels). The results show that \FW allows multiplexing more tasks than communication spatial streams. Finally, in Section~\ref{subsec:results_cifar10} we comprehensively evaluate \FW against baselines based on latency, energy consumption, communication and computation efficiency. The adaptation capability of \FW under variations in the communication environment is also evaluated and compared with MCR2/sm to demonstrate the effectiveness of the proposed techniques\textemdash namely, precoder/postcoder response and task-oriented pilot transmission\textemdash in addressing short-term and long-term changes, respectively. All results correspond to processing a full 64-element batch. In experiments involving \gls{mimo} computation-enabled methods, the metrics are normalized by the number of multiplexed tasks, providing average values per input. In the following, will use \textbf{SM} in tables and figures to indicate our \FW approach. \vspace{-0.2cm}

\subsection{Task Accuracy}\label{subsec:task_accuracy}

We multiplexed 4 tasks on 8 communication channels for image classification while 2 tasks were multiplexed on 8 communication channels for sentiment analysis. This setting allows a fair comparison of \FW with the baselines which do not provide multiplexing support for more tasks than physical streams. Note that full-stack baselines achieve identical accuracy in both \gls{los} and \gls{nlos}, as the use of a communication standard guarantees the exact representation transmitted from the mobile device is received at the edge server. Thus, their results are channel-independent and reported only once. In contrast, latent representations in methods MCR2/sm, PhyDNN/ss, and \FW are directly influenced by the wireless environment.

\smallskip
\noindent\textbf{Image classification.}
Table~\ref{tab:acc_CIFAR10} presents the classification accuracy considering the different versions of the WideResNet and using CIFAR-10. The evaluation on  CIFAR-100 and SVHN is reported in Appendix~\ref{subsec:results_other}. The results show that \FW delivers consistently high performance across all architectures and conditions. Specifically, it outperforms other approaches under both \gls{los} and \gls{nlos} conditions. The advantage over MCR2/sm stems from the joint optimization of communication and computation while MCR2/sm's linear precoder and postcoder are separately designed and thus suboptimal. Likewise, the performance gain over PhyDNN/ss is because the latter relies on a fixed codebook, which cannot adapt to varying channel conditions. \FW exhibits remarkable robustness, with the average performance drop of less than 4\% when transitioning from \gls{los} to \gls{nlos}, e.g., from 94.48\% to 90.92\% using WRN-28-10. This represents an improvement of 4.61\% and 4.16\% over MCR2/sm in the same conditions. \FW robustness is further evident in challenging scenarios --  even with the smallest architecture (WRN-10-4), \FW achieves 85.76\% accuracy in \gls{nlos}, surpassing the performance of MCR2/sm (85.36\%) and PhyDNN/ss (85.25\%) under favorable \gls{los} conditions by 0.40\% and 0.51\%, respectively. While full-stack methods achieve highest accuracy, it comes at the cost of increased latency as detailed in Section~\ref{subsec:results_cifar10}. In contrast, \FW attains comparable performance (93.69\% with WRN-16-8) with substantially lower latency.\vspace{-0.2cm}

\begin{table}[h]
  \small  
  \caption{Image classification. Experimental evaluation of \FW and baselines for different \glspl{dnn} on CIFAR-10. \vspace{-0.3cm}}
  \label{tab:acc_CIFAR10}
  \begingroup
  \setlength{\tabcolsep}{2pt}
  \renewcommand{\arraystretch}{0.85}  
  \begin{tabular}{@{}c|cccccc@{}}
    \toprule
    \textbf{WRN-28-10} & MNet & BF-9 & BF-3 & MCR2 & PhyDNN & \textbf{SM}\\
    \midrule
    \gls{los} & \multirow{2}{*}{94.74 \%} & \multirow{2}{*}{96.88 \%} & \multirow{2}{*}{96.42 \%} & 89.87 \% & 92.50 \% & 94.48 \%\\
    \gls{nlos} & & & & 86.76 \% & 88.23 \% & 90.92 \%\\
    \midrule
    \textbf{WRN-16-8} & MNet & BF-9 & BF-3 & MCR2 & PhyDNN & \textbf{SM}\\
    \midrule
    \gls{los} & \multirow{2}{*}{93.86 \%} & \multirow{2}{*}{95.96 \%} & \multirow{2}{*}{95.44 \%} & 89.17 \% & 91.85 \% & 93.69 \%\\
    \gls{nlos} & & & & 84.97 \% & 87.05 \% & 89.46 \%\\
    \midrule
    \textbf{WRN-10-4} & MNet & BF-9 & BF-3 & MCR2 & PhyDNN & \textbf{SM}\\
    \midrule
    \gls{los} & \multirow{2}{*}{87.71 \%} & \multirow{2}{*}{91.38 \%} & \multirow{2}{*}{90.13 \%} & 85.36 \% & 85.25 \% & 87.29 \%\\
    \gls{nlos} & & & & 83.40 \% & 83.34 \% & 85.76 \%\\
    \bottomrule
  \end{tabular}
  \endgroup
  \vspace{-0.1cm}
\end{table}

\noindent\textbf{Sentiment analysis.} Table~\ref{tab:acc_LRA} shows the results of the accuracy in the binary classification of sentiments using the Transformer architecture for text processing. The results follow the same trend discussed for image classification: \FW is the most effective approach among the \gls{phy} \gls{semcom} strategies and achieve accuracy values comparable to the ones of full protocol stack baselines.\vspace{-0.2cm}

\begin{table}[h]
  \small  
  \caption{Sentiment analysis.~Experimental evaluation of \FW and the baselines for the Transformer on \gls{lra}.\vspace{-0.3cm}}
  \label{tab:acc_LRA}
  \begingroup
  \setlength{\tabcolsep}{2pt}
  \renewcommand{\arraystretch}{0.85}  
  \begin{tabular}{@{}c|cccccc@{}}
    \toprule
    \textbf{Transformer} & MNet & BF-9 & BF-3 & MCR2 & PhyDNN & \textbf{SM}\\
    \midrule
    \gls{los} & \multirow{2}{*}{64.39 \%} & \multirow{2}{*}{66.06 \%} & \multirow{2}{*}{65.81 \%} & 60.93 \% & 61.04 \% & 64.02 \%\\
    \gls{nlos} & & & & 59.44 \% & 59.52 \% & 62.83 \%\\
    \bottomrule
  \end{tabular}
  \endgroup
  \vspace{-0.3cm}
\end{table}

\begin{figure*}[!t]
    \centering
    \begin{subfigure}{0.33\textwidth}
        \centering
        \includegraphics[width=\textwidth]{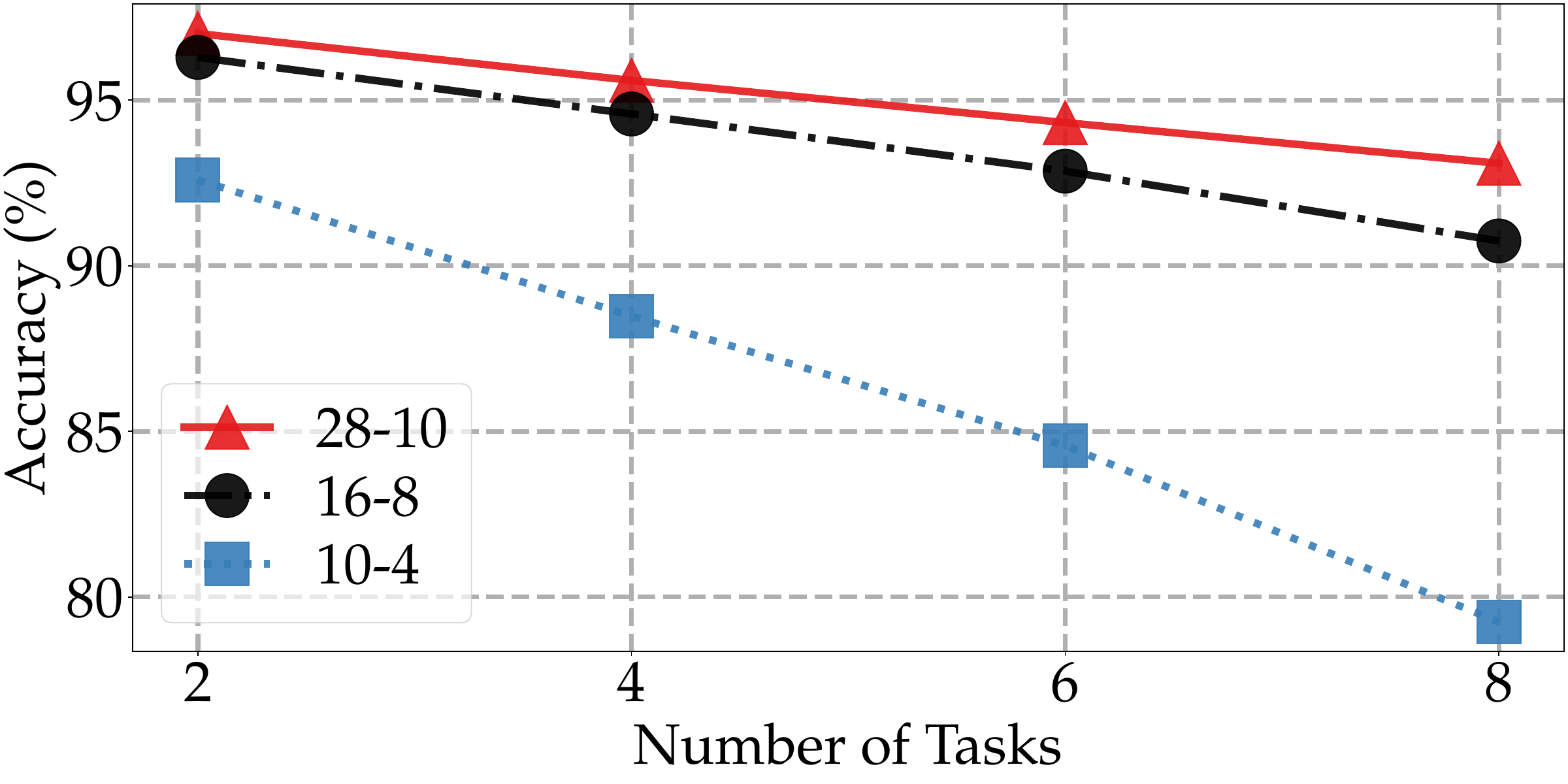}
        \caption{\footnotesize CIFAR-10, 4 comm. channels.}
        \label{fig:scalab_cifar10_4}
    \end{subfigure}
    \hfill
    \begin{subfigure}{0.33\textwidth}
        \centering
        \includegraphics[width=\textwidth]{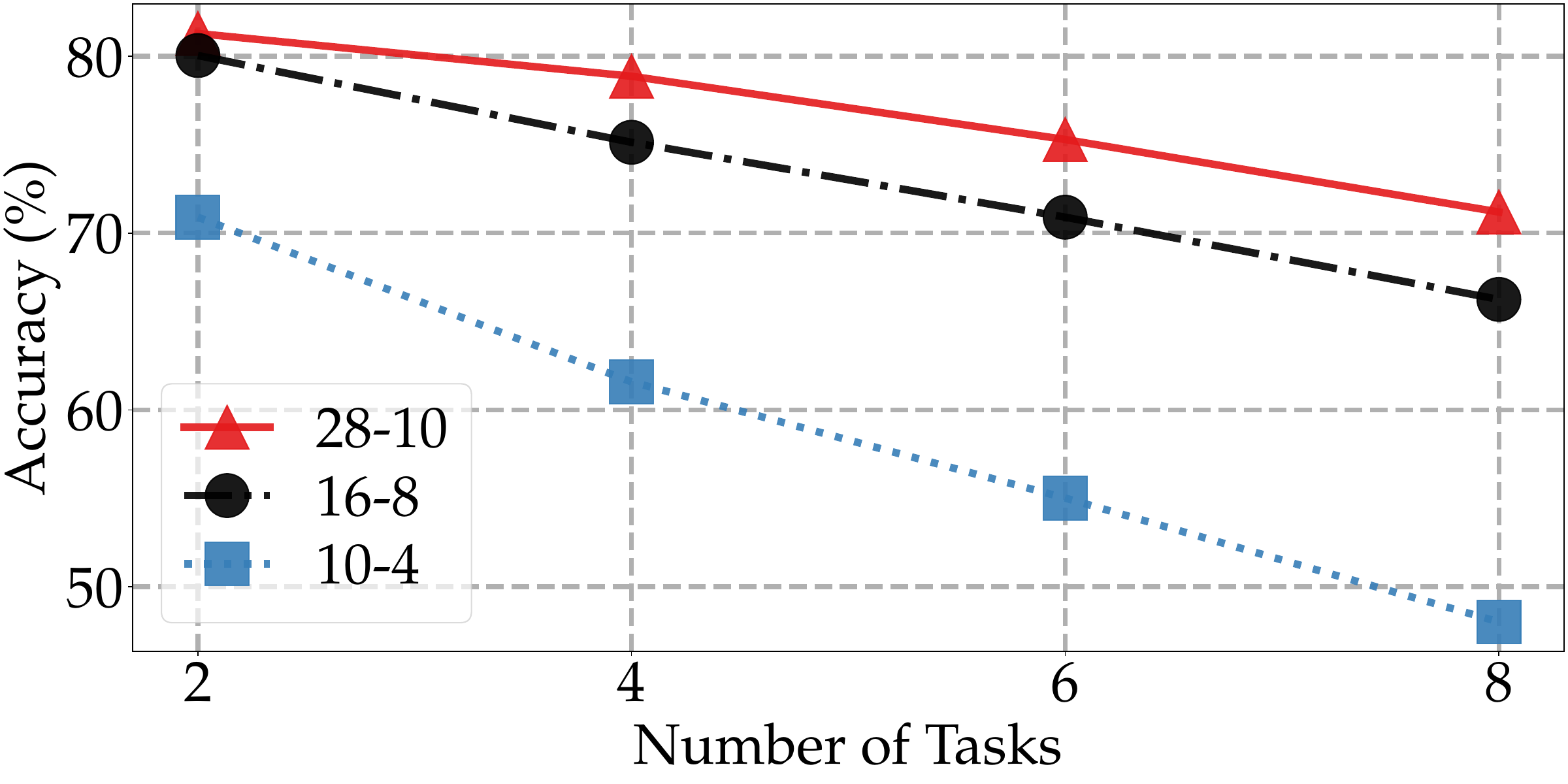}
        \caption{\footnotesize CIFAR-100, 4 comm. channels.}
        \label{fig:scalab_cifar100_4}
    \end{subfigure}
    \hfill
    \begin{subfigure}{0.33\textwidth}
        \centering
        \includegraphics[width=\textwidth]{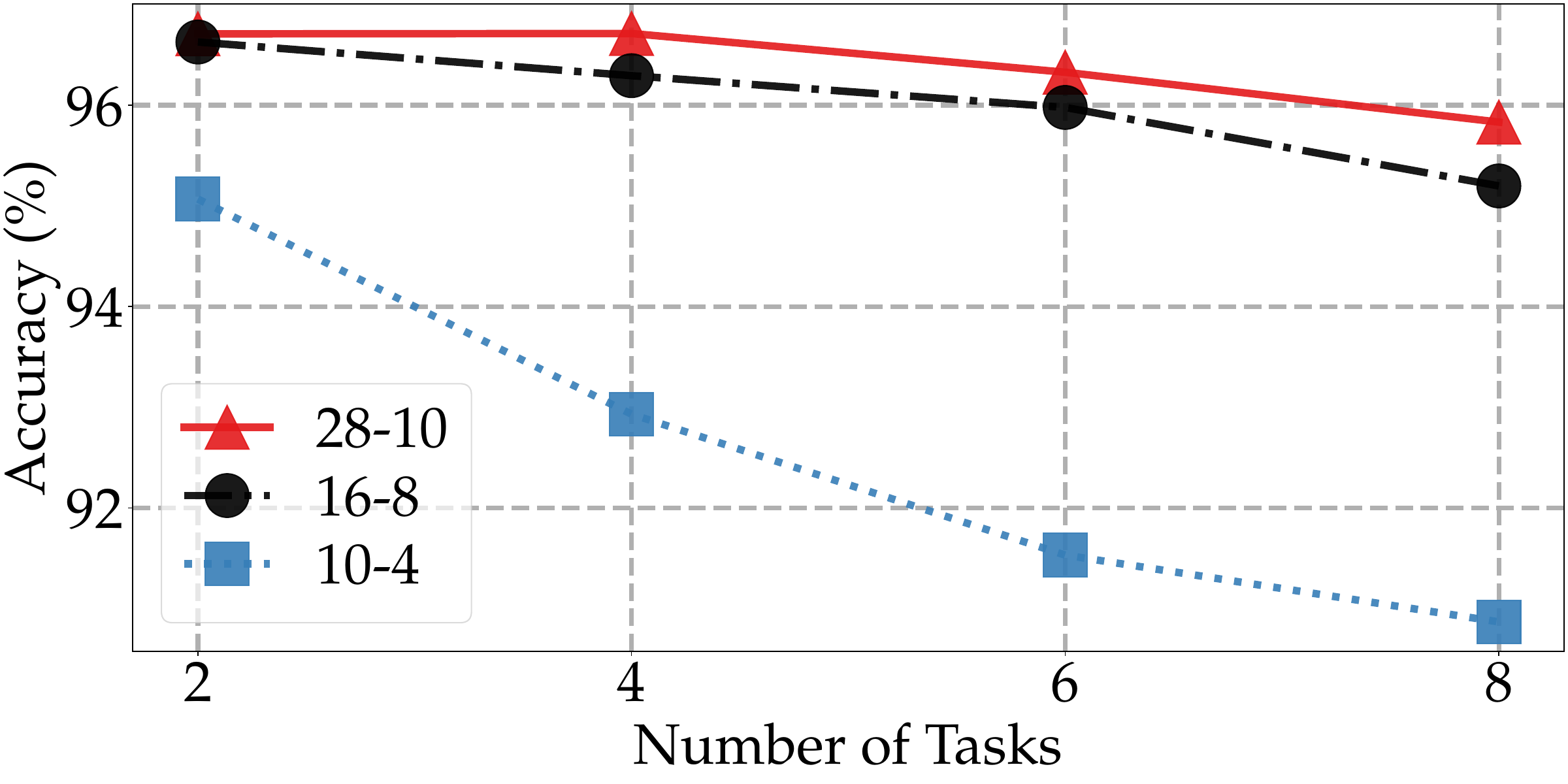}
        \caption{\footnotesize SVHN, 4 comm. channels.}
        \label{fig:scalab_svhn_4}
    \end{subfigure}
    \setlength\abovecaptionskip{-0.3cm}
    \caption{Image classification. Scalability evaluation with 2 and 4 communication channels and varying the number of multiplexed task inputs. `WRN-' is omitted in the legend for the different \glspl{dnn} for clarity.\vspace{-0.3cm}}
    \label{fig:scalability_image_class}
\end{figure*}

\begin{figure}[!t]
    \centering
    \includegraphics[width=0.7\linewidth]{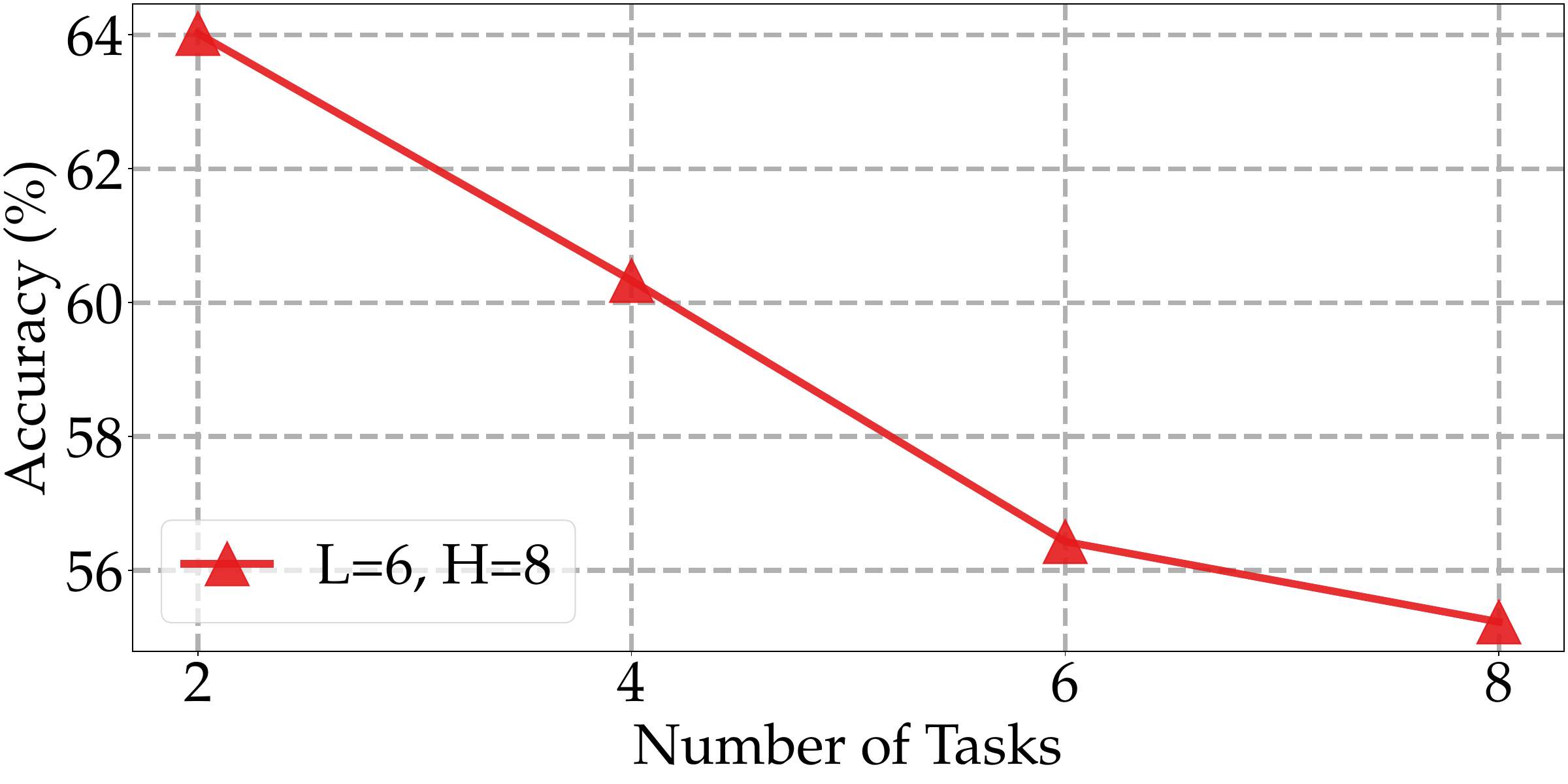}
    \setlength\abovecaptionskip{-0.01cm}
    \caption{Sentiment analysis. Scalability evaluation with 2 communication channels and varying the number of multiplexed task inputs.\vspace{-0.2cm}}
    \label{fig:scalability_sentiment}
\end{figure}

\subsection{Semantic Scalability}\label{subsec:scalability_evaluation}

We evaluate the scalability of \FW in terms of how the task accuracy varies based on the number of multiplexed tasks. 

Figure~\ref{fig:scalability_image_class} reports an extensive scalability evaluation performed using the image classification task and testing the performance of the different \glspl{dnn} on the three considered datasets. In general, the results demonstrate that \FW can multiplex more tasks than the number of physical streams supported by a given transceiver hardware, without noticeable degradation in individual task performance. More in detail, by comparing the results in the first three subplots -- obtained considering 4 communication streams and multiplexing 2, 4, 6 and 8 tasks at a time -- we notice that the performance drop is linked with the difficulty of the task (e.g., CIFAR-10 vs CIFAR-100) and the complexity of the \gls{dnn} used for processing (e.g., WRN-28-10 vs WRN-10-4). For example, Figure~\ref{fig:scalability_image_class}a shows that WRN-28-10 allows multiplexing 8 tasks guaranteeing a task accuracy of 93.09\% which is only 4\% less than the one achievable multiplexing 2 tasks (97.01\%), while reducing latency by 4$\times$ thanks to semantic multiplexing (8 streams are simultaneously processes instead of only 2). The same trend is visible in Figures~\ref{fig:scalability_image_class}b-c, with accuracy decrease rates that are dataset- and \gls{dnn}-specific. For example, for the more complex CIFAR-100 dataset (Figure~\ref{fig:scalability_image_class}b), accuracy drops by about 10\% when multiplexing 8 streams (71.19\%) instead of 2 (81.29\%) and its values are consistently smaller than the one achievable with the simpler 10-class datasets. 

Finally, in Figure~\ref{fig:scalability_sentiment}, we analyze \FW scalability on the sentiment analysis task which uses Transformers. The evaluation have been performed considering 2 communication channels. The results are promising also in this case, with less than 10\% accuracy drop when increasing the number of multiplexed tasks from 2 to 8.


\vspace{0.1cm}
\noindent\textbf{Disjoint Preprocessing.} In Table~\ref{tab:scale_over_disjoint_width}, we investigate the impact of the disjoint processing -- applied at the transmitter to the task inputs before binding (see Section~\ref{subsec:joint_processing}, Figure~\ref{fig:processing}) -- on the \FW scalability. Specifically, we compared the performance of \FW on image classification with CIFAR-10 when doubling the number of kernels in the convolutional layer used for the disjoint processing. The results are obtained using 4 communication channels while multiplexing 6 and 8 tasks and show that increasing the number of kernels ameliorates performance, especially when increasing the number of tasks to be multiplexed. This confirms that applying a preprocessing step before the binding helps enforcing orthogonality among the computation channels and can be optimized based on the specific task and hardware constraints.\vspace{-0.2cm}

\begin{table}[h]
  \small  
  \caption{Image classification. \FW accuracy for different \glspl{dnn} on CIFAR10 changing the number of kernels in the disjoint processing for 6 and 8 multiplexed tasks.\vspace{-0.3cm}}
  \label{tab:scale_over_disjoint_width}
  \begingroup
  \setlength{\tabcolsep}{2pt}
  \renewcommand{\arraystretch}{0.9}  
    \begin{tabular}{@{}c|cc@{}}
        \toprule
        \textbf{WRN-28-10} & 1$\times$ Kernels & 2$\times$ Kernels \\
        \midrule
        Tasks = 6 & 94.32 \% & 96.21 \%\\
        Tasks = 8 & 93.09 \% & 95.82 \%\\
        \midrule
        \textbf{WRN-16-8} & 1$\times$ Kernels & 2$\times$ Kernels \\
        \midrule
        Tasks = 6 & 92.85 \% & 93.93 \%\\
        Tasks = 8 & 90.74 \% & 91.82 \%\\
        \midrule
        \textbf{WRN-10-4} & 1$\times$ Kernels & 2$\times$ Kernels \\
        \midrule
        Tasks = 6 & 84.57 \% & 88.48 \%\\
        Tasks = 8 & 79.25 \% & 80.52 \%\\
        \bottomrule
    \end{tabular}
  \endgroup
 \vspace{-0.3cm}
\end{table}

\subsection{Resource Efficiency}\label{subsec:results_cifar10}

We evaluated \FW against the other baselines on the CIFAR-10 dataset for the computational analysis. We considered 8 communication channels and multiplexed 4 task inputs for this evaluation. We obtain the following figures of merit to evaluate the complexity (i) end-to-end execution time; (ii) energy consumption at the mobile device; (iii) total number of transmitted symbols from the mobile device to the edge server (communication efficiency); (iv) total number of \gls{dnn} parameters and operations performed by the mobile device (computation efficiency). For latency and energy measurements, evaluations were performed over 100 full executions (epochs) of the test set. To obtain reliable estimates of energy consumption, instantaneous power readings were sampled more than 1,000 times during data acquisition and averaged.

\begin{figure}[h]
    \begin{minipage}{0.485\linewidth}
    \centering
     \begin{subfigure}{\linewidth}
         \centering
         \includegraphics[width=\textwidth]{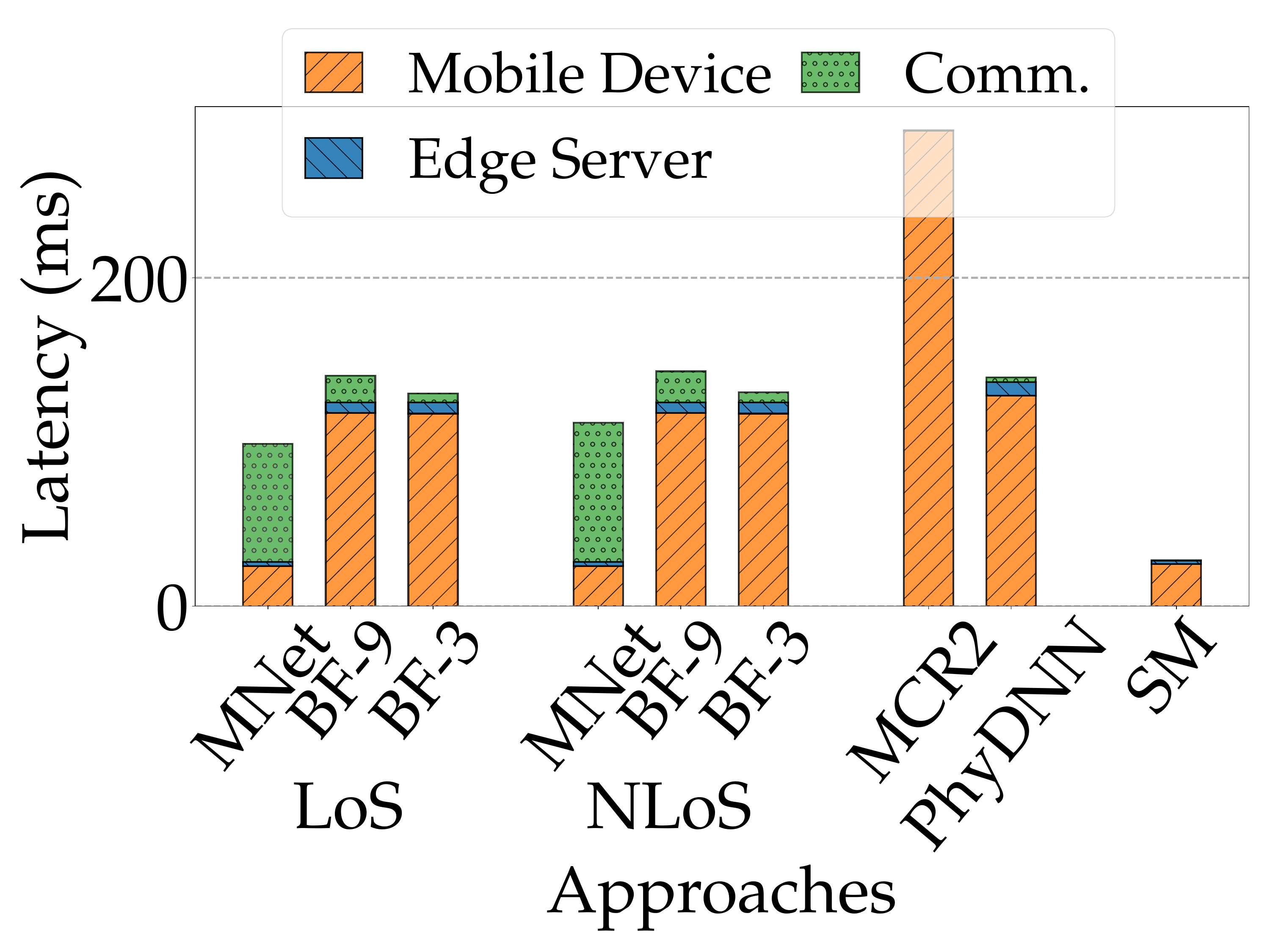}
         \caption{\footnotesize WRN-28-10.}
         \label{fig:latency_WRN-28-10}
     \end{subfigure}
     \begin{subfigure}{\linewidth}
         \centering
         \includegraphics[width=\textwidth]{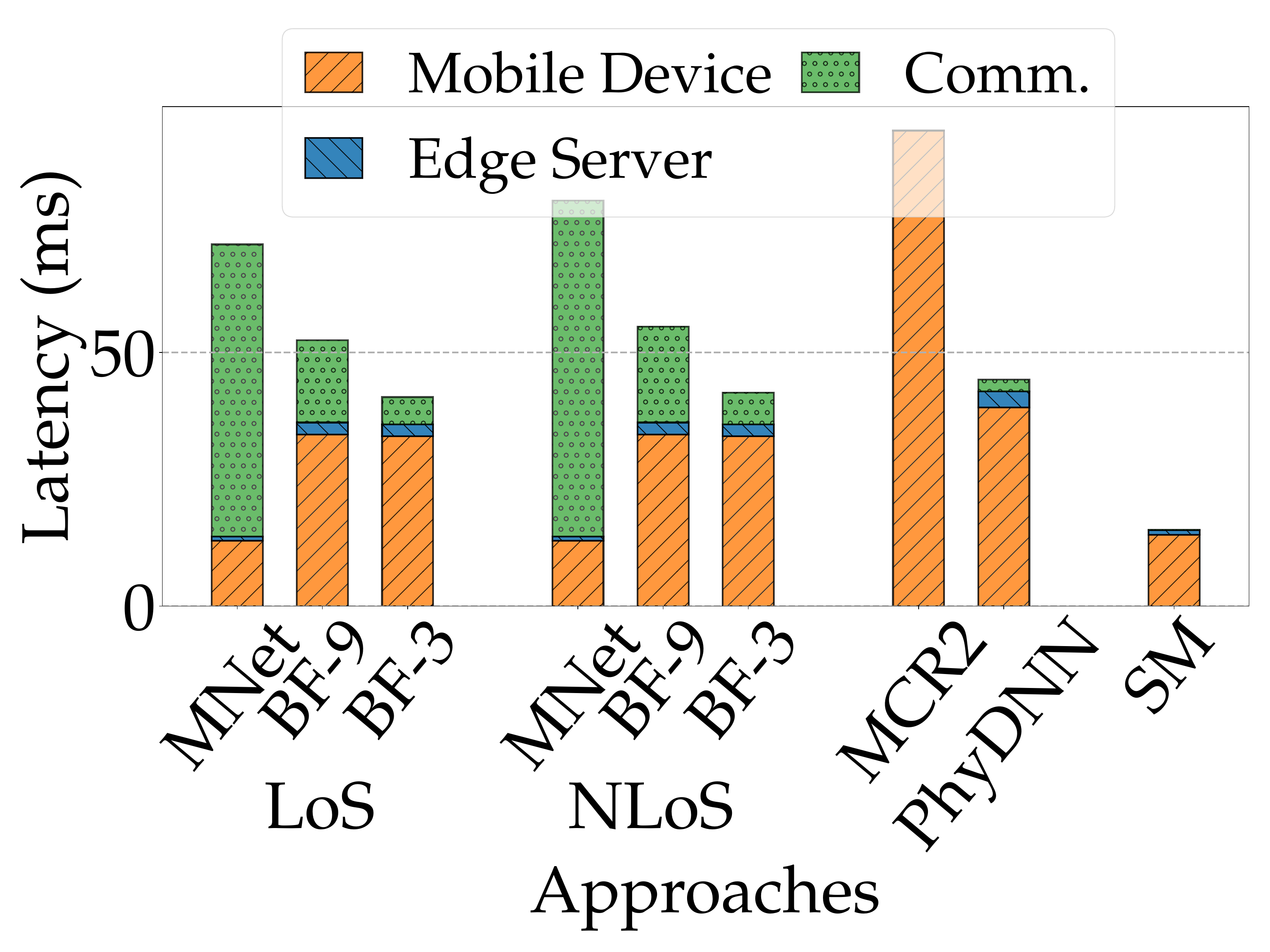}
         \caption{\footnotesize WRN-16-8.}
         \label{fig:latency_WRN-16-8}
     \end{subfigure}
     \begin{subfigure}{\linewidth}
         \centering
         \includegraphics[width=\textwidth]{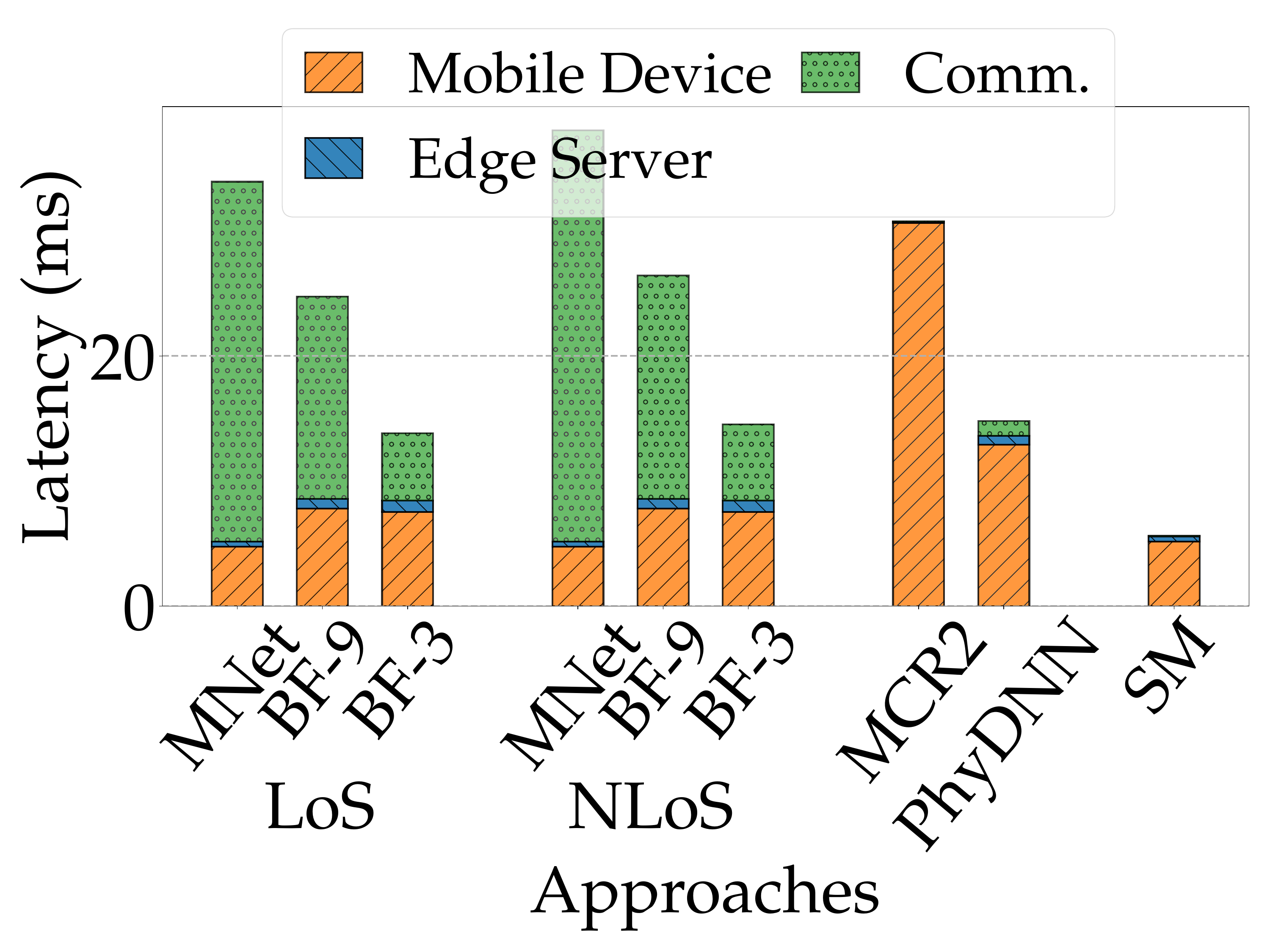}
         \caption{\footnotesize WRN-10-4.\vspace{-0.3cm}}
         \label{fig:latency_WRN-10-4}
     \end{subfigure}
     \caption{Breakdown of end-to-end latency.\vspace{-0.5cm}} 
     \label{fig:latency}
    \end{minipage}
    \hfill
    \begin{minipage}{0.485\linewidth}
    \centering
     \begin{subfigure}{\linewidth}
         \centering
         \includegraphics[width=\textwidth]{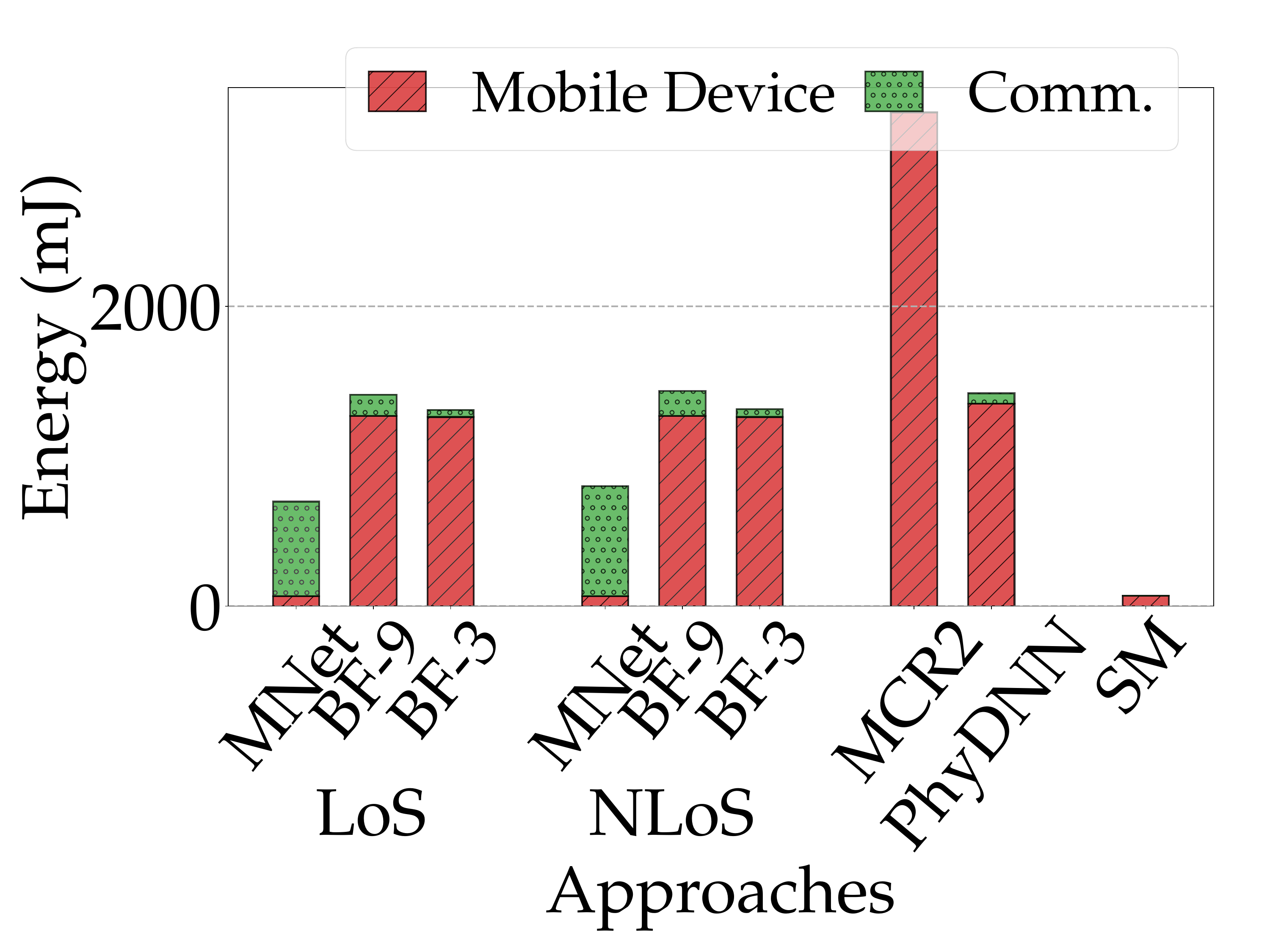}
         \caption{\footnotesize WRN-28-10.}
         \label{fig:energy_WRN-28-10}
     \end{subfigure}
     \begin{subfigure}{\linewidth}
         \centering
         \includegraphics[width=\textwidth]{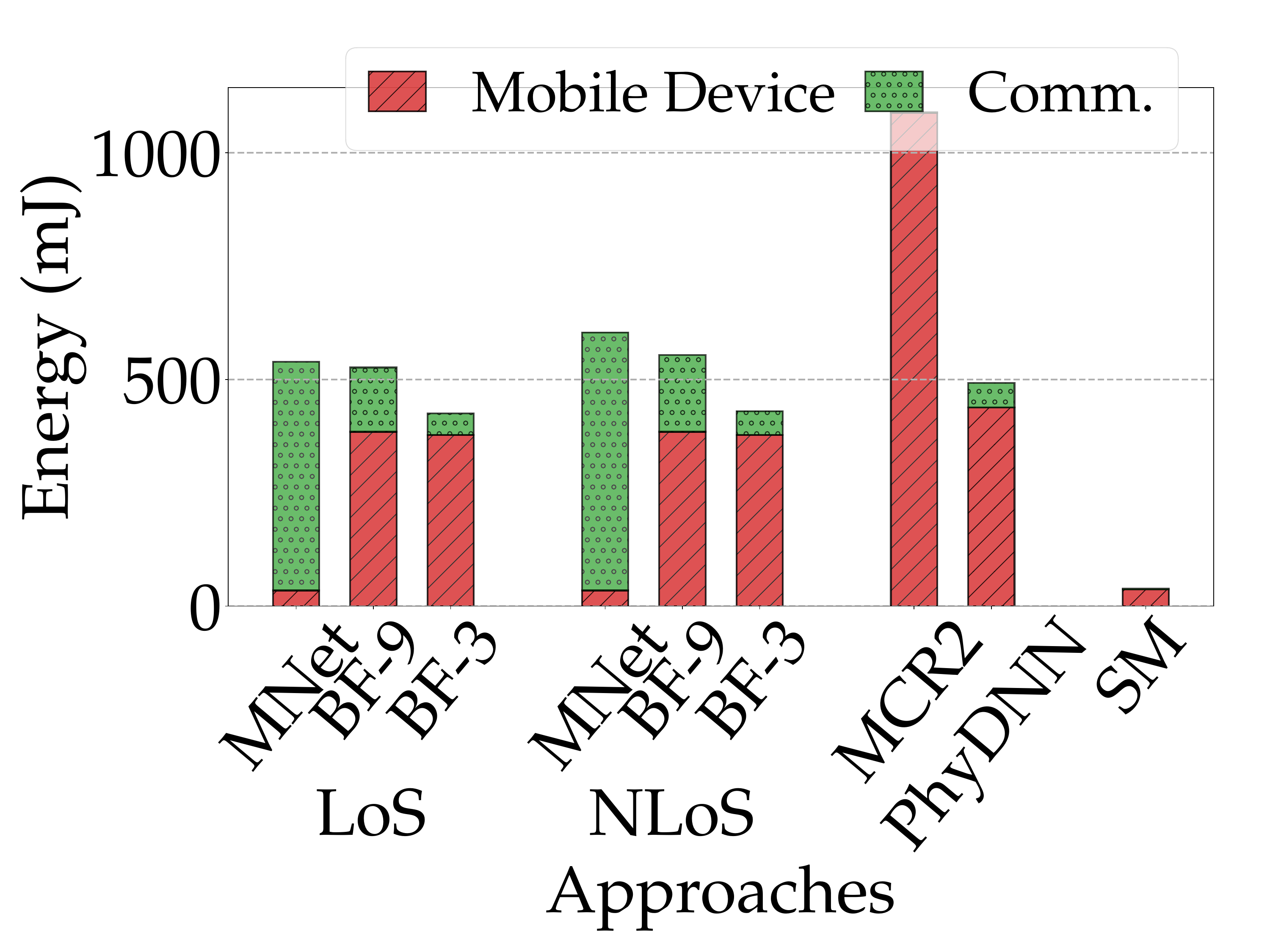}
         \caption{\footnotesize WRN-16-8.}
         \label{fig:energy_WRN-16-8}
     \end{subfigure}
     \begin{subfigure}{\linewidth}
         \centering
         \includegraphics[width=\textwidth]{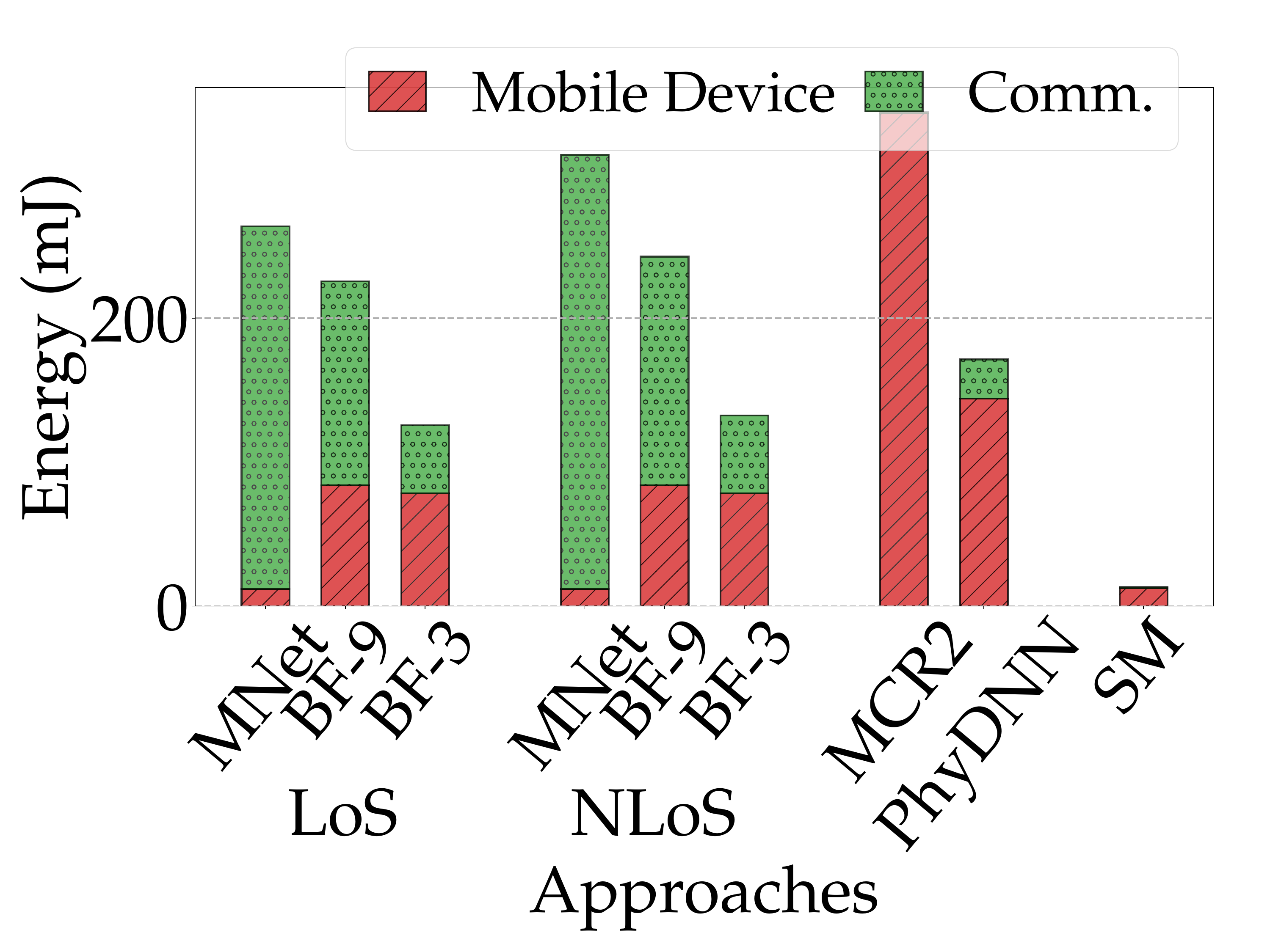}
         \caption{\footnotesize WRN-10-4.\vspace{-0.3cm}}
         \label{fig:energy_WRN-10-4}
     \end{subfigure}
     \caption{Mobile device energy consumption.\vspace{-0.5cm}} 
     \label{fig:energy}
    \end{minipage}
\end{figure}

\begin{figure*}[h]
    \centering
    \begin{minipage}{0.74\textwidth}
        \centering
        \begin{subfigure}{0.32\textwidth}
            \centering
            \includegraphics[width=\textwidth]{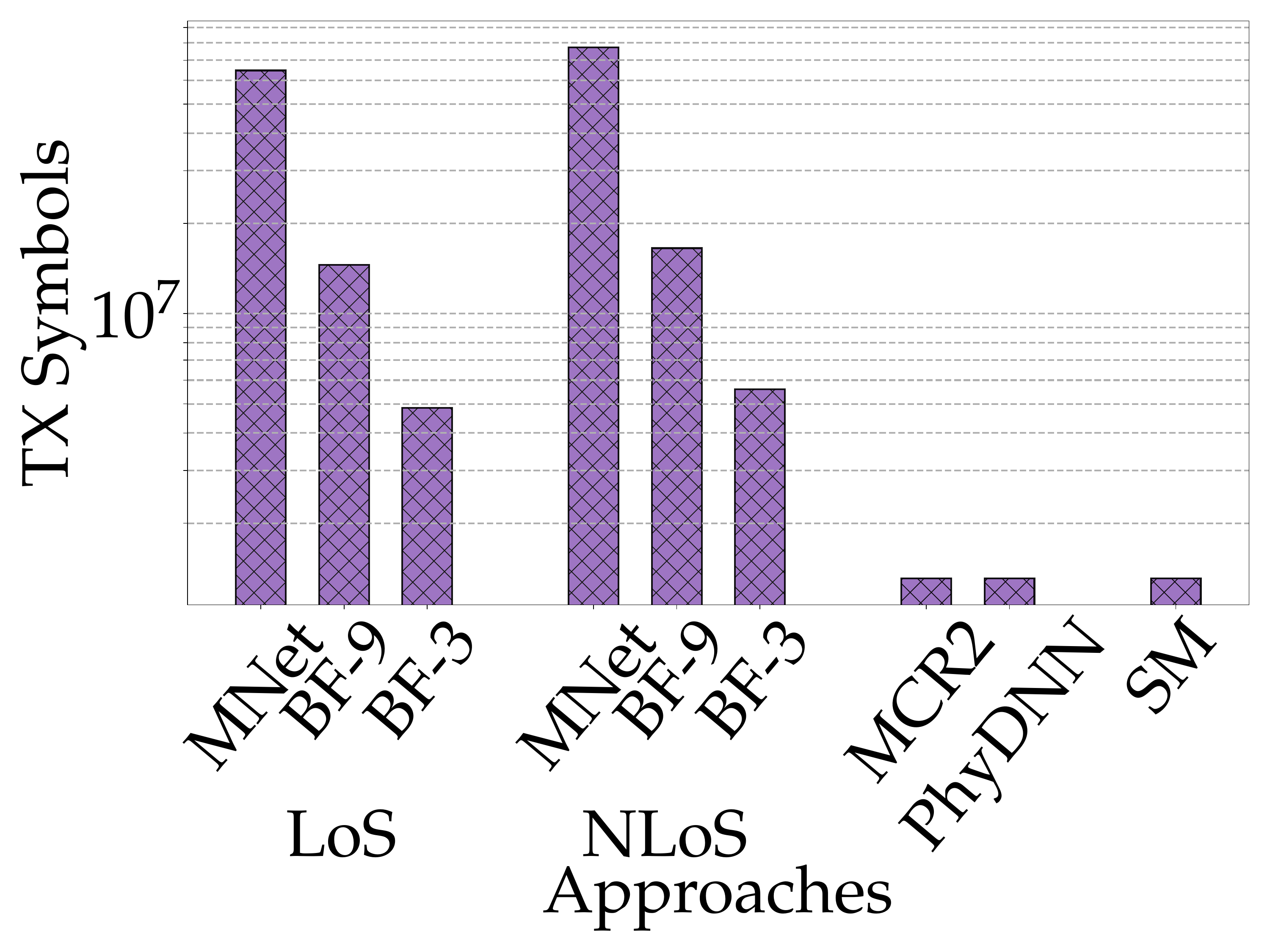}
            \caption{\footnotesize WRN-28-10.\vspace{-0.2cm}}
            \label{fig:symbols_WRN-28-10}
        \end{subfigure}
        \hfill
        \begin{subfigure}{0.32\textwidth}
            \centering
            \includegraphics[width=\textwidth]{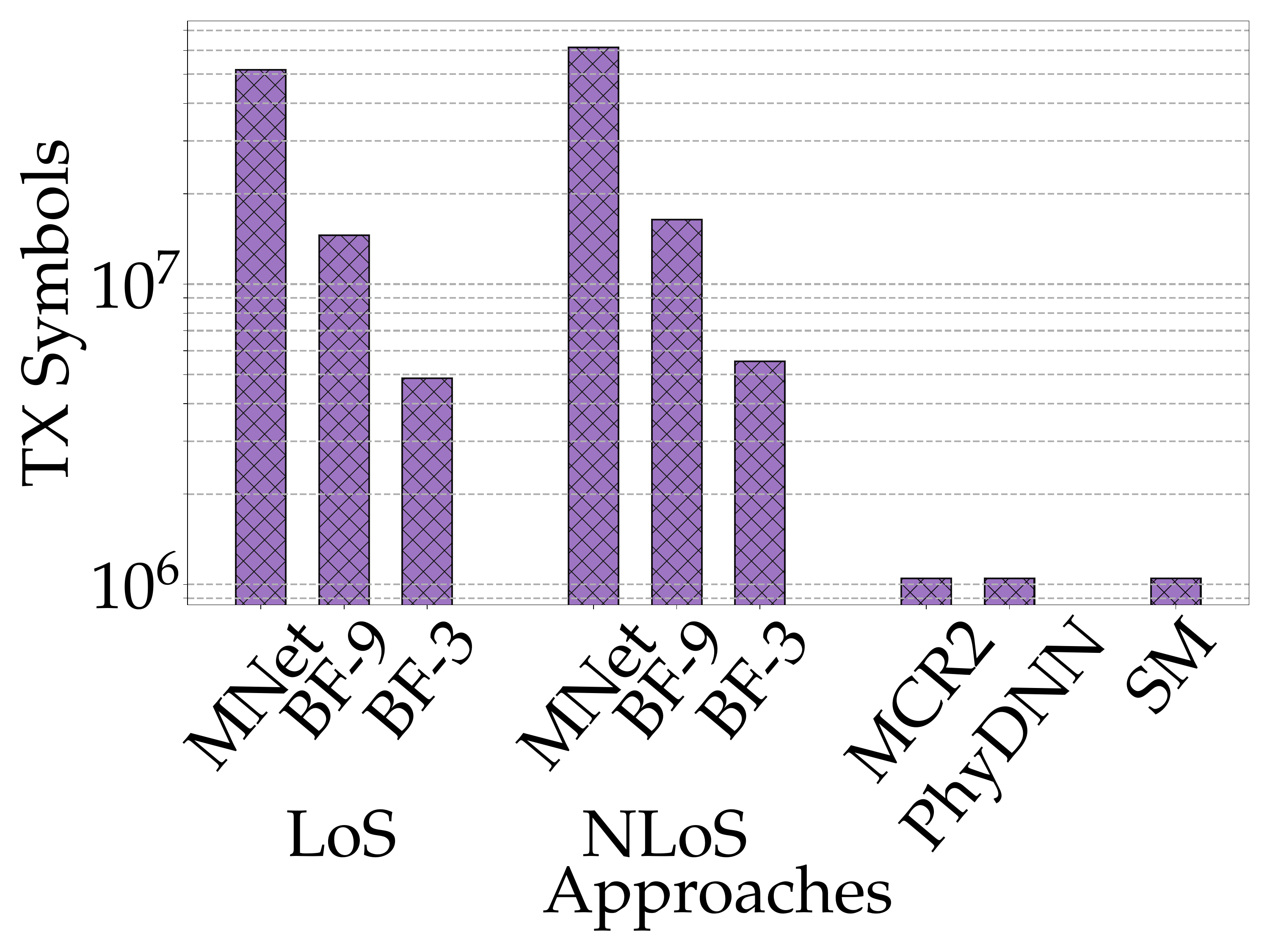}
            \caption{\footnotesize WRN-16-8.\vspace{-0.2cm}}
            \label{fig:symbols_WRN-16-8}
        \end{subfigure}
        \hfill
        \begin{subfigure}{0.32\textwidth}
            \centering
            \includegraphics[width=\textwidth]{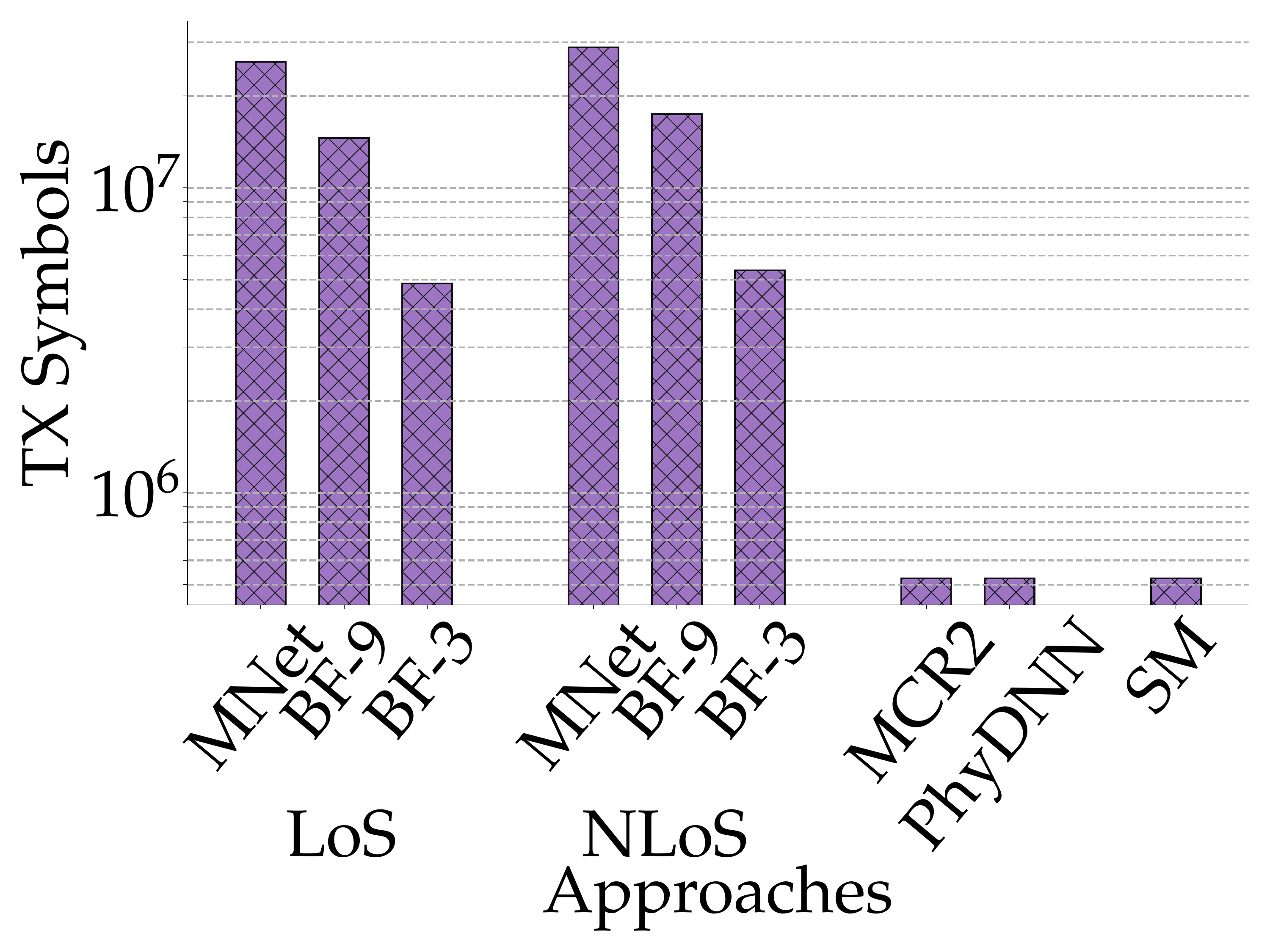}
            \caption{\footnotesize WRN-10-4.\vspace{-0.2cm}}
            \label{fig:symbols_WRN-10-4}
        \end{subfigure}
        \caption{Communication load comparison. The plots show the total number of transmitted \gls{i/q} symbols (payload and overhead combined) for different \glspl{dnn}.\vspace{-0.4cm}}
        \label{fig:efficiency}
    \end{minipage}
    \hfill
    \begin{minipage}{0.24\textwidth}
        \centering
        \scriptsize
        \captionof{table}{Memory and computational load of the head \gls{dnn} on the mobile device.\vspace{-0.3cm}} 
        \label{tab:comp}
        \begingroup
        \setlength{\tabcolsep}{2pt}
        \renewcommand{\arraystretch}{0.9}
        \begin{tabular}{@{}c|ccccc@{}}
            \toprule
             & MNet & BF & MCR2 & PhyDNN & \textbf{SM}\\
            \midrule
            \multicolumn{6}{c}{\textbf{WRN-28-10}}\\
            \midrule
            Params & 100 & 719 & 36474 & 1642 & 216\\
            FLOPs  & 47  & 875 & 5399  & 1830 & 92\\
            \midrule
            \multicolumn{6}{c}{\textbf{WRN-16-8}}\\
            \midrule
            Params & 66  & 328 & 10957 & 464  & 67\\
            FLOPs  & 27  & 389 & 1612  & 536  & 40\\
            \midrule
            \multicolumn{6}{c}{\textbf{WRN-10-4}}\\
            \midrule
            Params & 5   & 15  & 1795  & 48   & 5\\
            FLOPs  & 3   & 28  & 275   & 65   & 7\\
            \bottomrule
        \end{tabular}
        \endgroup
        \vspace{-0.3cm}
    \end{minipage}
\end{figure*}

\smallskip
\noindent \textbf{End-to-End Latency.}~The end-to-end latency accounts for (i) computation time on the mobile device (head \gls{dnn}); (ii) communication latency of sending latent representations; (iii) inference time on the edge server (tail \gls{dnn}). Figure~\ref{fig:latency_WRN-28-10}, \ref{fig:latency_WRN-16-8}, and \ref{fig:latency_WRN-10-4} report the breakdown of latency results for WRN-28-10, WRN-16-8, and WRN-10-4 architectures, respectively, under \gls{los} and \gls{nlos} scenarios. For methods employing superposition and precoding at the mobile device, and postcoding and decomposition at the edge server, the latency overhead of these modules is included in the reported inference times. Across all settings, \FW consistently shows lower latency compared to all baselines, including both full protocol stack and \gls{phy} approaches.

Communication latency is the primary bottleneck for full protocol stack approaches especially in \gls{nlos} scenarios, where frequent retransmissions further increase delay. By transmitting task-relevant features directly at the \gls{phy}, \FW avoids such an overhead. For instance, MNet/ms incurs up to 930$\times$ higher communication latency than \FW in \gls{nlos}. Although PhyDNN/ss also operates at the \gls{phy} and maintains consistent latency across channel conditions, its single-stream transmission limits efficiency. By leveraging semantic multiplexing, we achieve a 32$\times$ reduction in average latency with respect to PhyDNN/ss, since it can transmit semantic features in parallel across multiple streams. 

Beyond communication efficiency, \FW also achieves a notable reduction in computation latency compared to \gls{phy} baselines. Unlike MCR2/sm and PhyDNN/ss, \FW supports multi-task semantic processing, lowering computation latency by up to 11.3$\times$ and 5.0$\times$, respectively.

Overall, averaged across all backbones and channel conditions, \FW reduces the total latency by 4.46$\times$ over MNet/ms, 4.55$\times$ over BF-9/ss, 3.82$\times$ over BF-3/ss, 8.53$\times$ over MCR2/sm, and 4.09$\times$ over PhyDNN/ss.

\vspace{0.1cm}
\noindent \textbf{Energy Consumption.}~We evaluate the energy consumption, including (i) energy consumption during transmitter-side \gls{dnn} execution; (ii) energy consumed to communicate the latent. As illustrated in Figure~\ref{fig:energy_WRN-28-10}, \ref{fig:energy_WRN-16-8}, and \ref{fig:energy_WRN-10-4}, the overall energy footprint varies significantly with the \gls{dnn} size and the communication environment. Full-stack baselines incur high communication energy, especially in \gls{nlos}, while MCR2/sm shows high computation energy due to its heavy processing on the mobile device.

In contrast, \FW maintains a consistently low energy profile across all architectures. It achieves up to 187.51$\times$ lower communication energy than full-stack methods on average with WRN-10-4, and reduces computation energy by 33.97$\times$ compared to task-oriented \gls{semcom} approaches on average with WRN-28-10. Even in the most demanding case (WRN-28-10 under \gls{nlos}), \FW remains significantly more efficient, with a total energy usage well below all baselines (11.36$\times$ lower than MNet/ms which is the second best method). This efficiency stems from \FW's strategic computation distribution, where only the first layer and the lightweight precoding are executed on the mobile device, while the energy-intensive deeper layers are offloaded to the edge server. Furthermore, its efficiency gain arises from task multiplexing, which reduces the average energy consumption per input.

\smallskip
\noindent \textbf{Communication and Computation Efficiency.}~The total number of transmitted \gls{i/q} symbols—including both payload and protocol overhead, measured for various approaches under both \gls{los} and \gls{nlos} conditions, is depicted in Figures~\ref{fig:efficiency}a--\ref{fig:efficiency}c. Full protocol stack baselines exhibit the highest communication overhead, while \FW maintains a near-constant transmission volume of just 1.3 $\times 10^6$ symbols regardless of the environment by bypassing all the overhead introduced by the different layers of the stack. On average, \FW reduces the number of transmitted symbols by 54.24$\times$ compared to MNet/ms, 16.73$\times$ compared to BF-9/ss, and 5.56$\times$ compared to BF-3/ss across \gls{los}/\gls{nlos} scenarios and models.

Table~\ref{tab:comp} reports the number of parameters and \glspl{flop} required to execute the head \glspl{dnn} across all evaluated methods and WideResNet architectures. While MCR2/sm exhibits the highest computational load at the transmitter—reaching over 36~MB and 5.3 billion MACs in WRN-28-10—other full-stack baselines also impose non-trivial memory and compute overhead.
\FW guarantees high task performance (as shown in Table~\ref{tab:acc_CIFAR10}) with a substantially lower computational and memory footprint. For instance, in WRN-10-4, it requires 42.9~KB of memory and 13.2~MMac, which is significantly lower than all baselines. Even in deeper architectures like WRN-28-10, \FW keeps its computation below 100 MMac and memory footprint under 0.5 MB.

\begin{figure}[h]
    \centering
    \includegraphics[width=\linewidth]{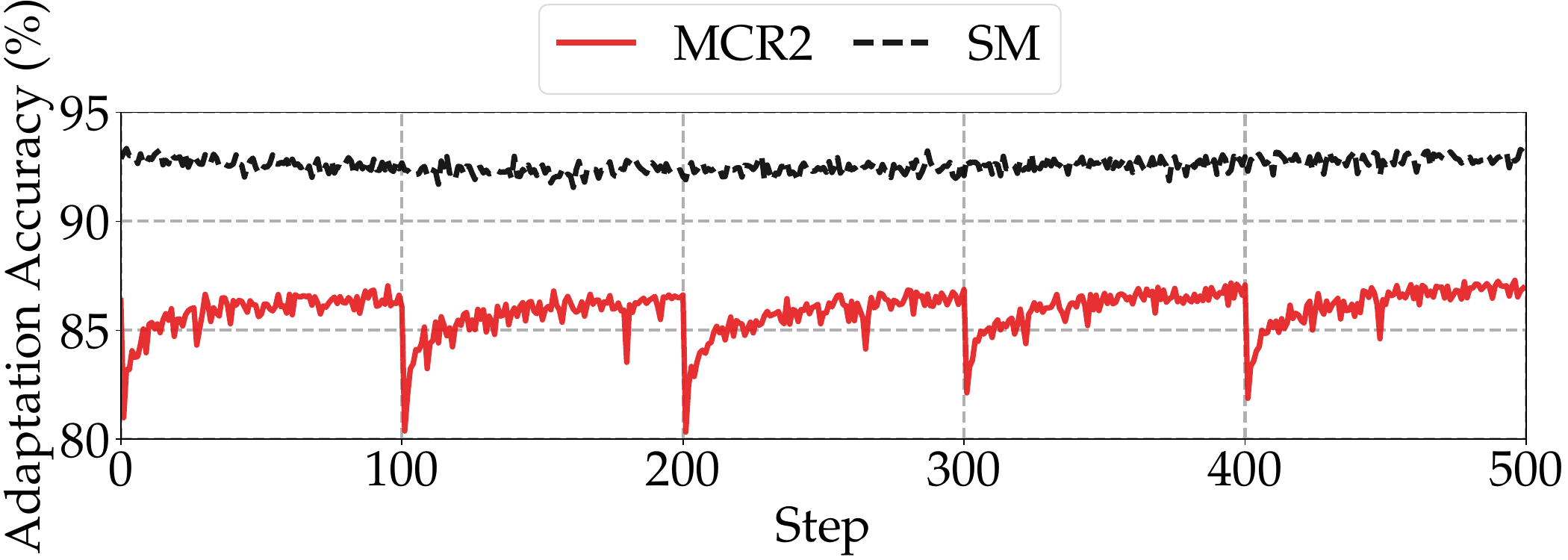}
    \setlength\abovecaptionskip{-0.2cm}
    \caption{Comparison between \FW and MCR2/sm in inference-time adaptation to varying channels.\vspace{-0.3cm}}
    \label{fig:adapt}
\end{figure}

\smallskip
\noindent \textbf{Adaptation to Dynamic Channels.}~We simulated a dynamic wireless channel by changing the positions of the transmitter and receiver every 100 inference steps, resulting in new \gls{csi} realizations. Classification accuracy on CIFAR-10 using WRN-28-10 is recorded continuously to assess each method's ability to maintain task performance under these variations. As shown in Figure~\ref{fig:adapt}, \FW maintains consistently high accuracy throughout the adaptation sequence, while MCR2/sm exhibits frequent and severe drops in performance. This discrepancy stems from the fact that MCR2/sm depends on an iterative precoding algorithm that solves an optimization problem for each new \gls{csi} snapshot. However, this approach introduces instability, especially when the channel varies quickly, and leads to significant task performance degradation. In contrast, \FW employs a stochastic, learnable precoder to handle dynamic channel conditions. By incorporating randomness into its design, the precoder compensate for the channel distortions and enables generalization across diverse environments ensuring stable performance without runtime overhead. Quantitatively, \FW achieves an average classification accuracy of 92.56\% over 500 adaptation steps, with a minimum of 91.55\% and a maximum of 93.33\%. In contrast, MCR2/sm accuracy is on average only 85.88\%, with drops down to 80.33\%.\vspace{-0.2cm}

\section{Related Work}\label{sec:related-work}

While the goal of the first \gls{semcom} systems was the exact recovery of the transmitter (input) message or its semantics at the receiver (output), later studies replaced the reconstruction loss with a task loss to achieve task-oriented \gls{semcom}~\cite{xie2023robust, shao2021learning}. The authors in~\cite{jankowski2020wireless} proposed a \gls{jscc} scheme by using an autoencoder architecture with a noise-injection layer that models the wireless channel, and train it for an image retrieval task. In~\cite{abdi2025phydnns}, the authors propose a task-oriented \gls{semcom} system that generates discrete latent symbols to be directly modulated and transmitted as a waveform. They take an already-trained \gls{dnn} and fine-tune it considering the channel distortions so that it can be brought to the \gls{phy} layer. However, the proposed system only supports \gls{siso} communication and fails to exploit the high communication rates enabled by \gls{mimo} communication systems. The utilization of \gls{mimo} transceivers has been investigated in~\cite{wu2024deep, xie2022task, zhang2024scan} by directly reusing traditional \gls{mimo} techniques. However, such designs focus on communication metrics such as throughput maximization, \gls{mse} and \gls{ber} minimization, which are inconsistent with task performance goals~\cite{wen2023task}. \Citet{cai2025end,cai2024multi} used maximal coding rate reduction as a substitute objective for the end task performance to train the \gls{mimo} system. They proposed a \gls{dnn}-based algorithm to optimize the linear precoding and postcoding blocks adopted in their system. However, they optimize the wireless communication and the \gls{dnn}-based codecs separately. Such design cannot achieve an end-to-end optimal system where both communication and computation are learned with the aim of successful task completion. Moreover, approaches in the literature are unable to multiplex computation and transmission of more computing tasks than the number of physically available channels.\vspace{-0.1cm}

\section{Conclusions and Next Steps}\label{sec:conclusion}

In this work, we have proposed \textit{Semantic Multiplexing}, a new \gls{semcom} strategy to jointly process and transmit several computing tasks at the semantic layer. \FW addresses the key bottleneck of current approaches being the separate optimization of the communication and computation pipelines which notoriously requires a complicated design. For this, we analytically formulate a probabilistic model of the transmitter and receiver system and trained it in a task-oriented fashion. A precoder and postcoder are designed to allow the transmitter and receiver to compensate for the wireless channel impairments. To enable the joint optimization of the communication and computation objectives, we developed and integrated a wireless channel model into the system. We developed a semantic channel sounding strategy on top of the standard sounding to make the \FW modules adapt to the varying channel conditions.  We prototyped \FW on an experimental and extensively evaluated its performance on image classification and sentiment analysis.\vspace{0.1cm}

Our experimental results have shown that \FW outperforms state-of-the-art \gls{semcom} approaches allowing the system to semantically multiplex more tasks than the number of physical streams. We believe our study may open several exciting research avenues in the field of semantic networking. For example, an intriguing direction would be investigating the tradeoffs at the basis of semantic multiplexing and its optimization depending on the specific applications. We have experimentally shown in Section~\ref{subsec:scalability_evaluation} that multiplexing performance changes based on the specific task and the approach to address it (semantic processing). We also revealed that the design of the disjoint processing before binding the input tasks also impacts the performance. While in our study we started evaluating these tradeoff between performance and computational complexity experimentally, future research may target a further analytical analysis of these design parameters to investigate these tradeoff at a fundamental level. Indeed, the upper bound on the number of tasks that can be multiplexed based on the problem constraints is an open and intriguing research question trigger by the new semantic communication approach we proposed in this paper.\vspace{-0.1cm}

\bibliographystyle{ACM-Reference-Format}
\bibliography{ref/bibliography, ref/francesco}

@misc{AppleVisionPro_ProductPage_2025,
  author       = {{Apple Inc.}},
  title        = {Apple Vision Pro},
  howpublished = {Product page},
  year         = {2025},
  url          = {https://www.apple.com/apple-vision-pro/},
  note         = {Accessed: 2025-08-30}
}

@article{liang2023dnn,
  title={DNN surgery: Accelerating DNN inference on the edge through layer partitioning},
  author={Liang, Huanghuang and Sang, Qianlong and Hu, Chuang and Cheng, Dazhao and Zhou, Xiaobo and Wang, Dan and Bao, Wei and Wang, Yu},
  journal={IEEE transactions on Cloud Computing},
  volume={11},
  number={3},
  pages={3111--3125},
  year={2023},
  publisher={IEEE}
}

@misc{kingma2013auto,
  title={Auto-encoding variational bayes},
  author={Kingma, Diederik P and Welling, Max and others},
  year={2013},
  publisher={Banff, Canada}
}

@article{murahari2023mux,
  title={{Mux-plms: Data multiplexing for high-throughput language models}},
  author={Murahari, Vishvak and Deshpande, Ameet and Jimenez, Carlos E and Shafran, Izhak and Wang, Mingqiu and Cao, Yuan and Narasimhan, Karthik},
  journal={arXiv preprint arXiv:2302.12441},
  year={2023}
}

@article{kleyko2022survey,
  title={A survey on hyperdimensional computing aka vector symbolic architectures, part i: Models and data transformations},
  author={Kleyko, Denis and Rachkovskij, Dmitri A and Osipov, Evgeny and Rahimi, Abbas},
  journal={ACM Computing Surveys},
  volume={55},
  number={6},
  pages={1--40},
  year={2022},
  publisher={ACM New York, NY}
}

@inproceedings{maas-etal-2011-learning,
    title = {{Learning Word Vectors for Sentiment Analysis}},
    author = "Maas, Andrew L.  and
      Daly, Raymond E.  and
      Pham, Peter T.  and
      Huang, Dan  and
      Ng, Andrew Y.  and
      Potts, Christopher",
    editor = "Lin, Dekang  and
      Matsumoto, Yuji  and
      Mihalcea, Rada",
    booktitle = "Proceedings of the 49th Annual Meeting of the Association for Computational Linguistics: Human Language Technologies",
    month = jun,
    year = "2011",
    address = "Portland, Oregon, USA",
    publisher = "Association for Computational Linguistics",
    url = "https://aclanthology.org/P11-1015/",
    pages = "142--150"
}

@inproceedings{tay2021long,
title={{Long Range Arena : A Benchmark for Efficient Transformers}},
author={Yi Tay and Mostafa Dehghani and Samira Abnar and Yikang Shen and Dara Bahri and Philip Pham and Jinfeng Rao and Liu Yang and Sebastian Ruder and Donald Metzler},
booktitle={International Conference on Learning Representations},
year={2021},
url={https://openreview.net/forum?id=qVyeW-grC2k}
}

@article{jermyn2005invariant,
  title={{Invariant Bayesian estimation on manifolds}},
  author={Jermyn, Ian H},
  year={2005}
}

@article{jang2022reparametrization,
  title={{A reparametrization-invariant sharpness measure based on information geometry}},
  author={Jang, Cheongjae and Lee, Sungyoon and Park, Frank and Noh, Yung-Kyun},
  journal={Advances in neural information processing systems},
  volume={35},
  pages={27893--27905},
  year={2022}
}

@article{geraci2025wi,
	author = {Geraci, Giovanni and Meneghello, Francesca and Wilhelmi, Francesc and Lopez-Perez, David and Val, I{\~n}aki and Giordano, Lorenzo Galati and Cordeiro, Carlos and Ghosh, Monisha and Knightly, Edward and Bellalta, Boris},
	journal = {arXiv preprint arXiv:2507.09613},
	title = {{Wi-Fi: Twenty-Five Years and Counting}},
	year = {2025}}

@article{gallant2013representing,
	author = {Gallant, Stephen I and Okaywe, T Wendy},
	journal = {Neural Computation},
	number = {8},
	pages = {2038--2078},
	publisher = {MIT Press},
	title = {{Representing Objects, Relations, and Sequences}},
	volume = {25},
	year = {2013}}

@article{alemi2016deep,
	author = {Alemi, Alexander A and Fischer, Ian and Dillon, Joshua V and Murphy, Kevin},
	journal = {arXiv preprint arXiv:1612.00410},
	title = {{Deep Variational Information Bottleneck}},
	year = {2016}}

@article{plate1995holographic,
	author = {Plate, Tony A},
	journal = {IEEE Transactions on Neural networks},
	number = {3},
	pages = {623--641},
	publisher = {IEEE},
	title = {{Holographic Reduced Representations}},
	volume = {6},
	year = {1995}}

@article{murahari2022datamux,
	author = {Murahari, Vishvak and Jimenez, Carlos and Yang, Runzhe and Narasimhan, Karthik},
	journal = {Advances in Neural Information Processing Systems},
	pages = {17515--17527},
	title = {{DataMUX: Data Multiplexing for Neural Networks}},
	volume = {35},
	year = {2022}}

@article{menet2023mimonets,
	author = {Menet, Nicolas and Hersche, Michael and Karunaratne, Geethan and Benini, Luca and Sebastian, Abu and Rahimi, Abbas},
	journal = {Advances in Neural Information Processing Systems},
	pages = {39553--39565},
	title = {{MIMONets: Multiple-Input-Multiple-Output Neural Networks Exploiting Computation in Superposition}},
	volume = {36},
	year = {2023}}

@article{wen2023task,
	author = {Wen, Dingzhu and Jiao, Xiang and Liu, Peixi and Zhu, Guangxu and Shi, Yuanming and Huang, Kaibin},
	journal = {IEEE Transactions on Wireless Communications},
	number = {3},
	pages = {2039--2053},
	publisher = {IEEE},
	title = {{Task-Oriented Over-the-Air Computation for Multi-Device Edge AI}},
	volume = {23},
	year = {2023}}

@article{cai2025end,
	author = {Cai, Chang and Yuan, Xiaojun and Zhang, Ying-Jun Angela},
	journal = {IEEE Journal on Selected Areas in Communications},
	publisher = {IEEE},
	title = {End-to-End Learning for Task-Oriented Semantic Communications Over MIMO Channels: An Information-Theoretic Framework},
	year = {2025}}

@article{shao2021learning,
	author = {Shao, Jiawei and Mao, Yuyi and Zhang, Jun},
	journal = {IEEE Journal on Selected Areas in Communications},
	number = {1},
	pages = {197--211},
	publisher = {IEEE},
	title = {{Learning Task-Oriented Communication for Edge Inference: An Information Bottleneck Approach}},
	volume = {40},
	year = {2021}}

@inproceedings{zagoruyko2016wide,
	author = {Zagoruyko, Sergey and Komodakis, Nikos},
	booktitle = {British Machine Vision Conference 2016},
	organization = {British Machine Vision Association},
	title = {{Wide Residual Networks}},
	year = {2016}}

@article{cai2024multi,
	author = {Cai, Chang and Yuan, Xiaojun and Angela Zhang, Ying-Jun},
	doi = {10.1109/TWC.2024.3461336},
	journal = {IEEE Transactions on Wireless Communications},
	number = {12},
	pages = {18096-18110},
	title = {{Multi-Device Task-Oriented Communication via Maximal Coding Rate Reduction}},
	volume = {23},
	year = {2024}}

@article{abdi2025phydnns,
	author = {Mohammad Abdi and Khandaker Foysal Haque and Francesca Meneghello and Jonathan Ashdown and Francesco Restuccia},
	journal = {Proceedings of IEEE International Conference on Computer Communications (INFOCOM)},
	title = {{PhyDNNs: Bringing Deep Neural Networks to the Physical Layer}},
	year = 2025}

@article{xie2023robust,
	author = {Xie, Songjie and Ma, Shuai and Ding, Ming and Shi, Yuanming and Tang, Mingjian and Wu, Youlong},
	journal = {IEEE Journal on Selected Areas in Communications},
	number = {8},
	pages = {2577--2591},
	publisher = {IEEE},
	title = {{Robust Information Bottleneck for Task-Oriented Communication with Digital Modulation}},
	volume = {41},
	year = {2023}}

@article{jankowski2020wireless,
	author = {Jankowski, Mikolaj and G{\"u}nd{\"u}z, Deniz and Mikolajczyk, Krystian},
	journal = {IEEE Journal on Selected Areas in Communications},
	number = {1},
	pages = {89--100},
	publisher = {IEEE},
	title = {{Wireless Image Retrieval at the Edge}},
	volume = {39},
	year = {2020}}

@article{blei2017variational,
	author = {Blei, David M and Kucukelbir, Alp and McAuliffe, Jon D},
	journal = {Journal of the American statistical Association},
	number = {518},
	pages = {859--877},
	publisher = {Taylor \& Francis},
	title = {{Variational Inference: A Review for Statisticians}},
	volume = {112},
	year = {2017}}

@article{wu2024deep,
	author = {Wu, Haotian and Shao, Yulin and Bian, Chenghong and Mikolajczyk, Krystian and G{\"u}nd{\"u}z, Deniz},
	journal = {IEEE Transactions on Wireless Communications},
	month = {October},
	number = {10},
	pages = {15002--15017},
	publisher = {IEEE},
	title = {{Deep Joint Source-Channel Coding for Adaptive Image Transmission Over MIMO Channels}},
	volume = {23},
	year = {2024}}

@article{zhang2024scan,
	author = {Zhang, Guangyi and Hu, Qiyu and Cai, Yunlong and Yu, Guanding},
	journal = {IEEE Transactions on Cognitive Communications and Networking},
	month = {October},
	number = {5},
	pages = {1759--1773},
	publisher = {IEEE},
	title = {{SCAN: Semantic Communication with Adaptive Channel Feedback}},
	volume = {10},
	year = {2024}}

@article{xie2022task,
	author = {Xie, Huiqiang and Qin, Zhijin and Tao, Xiaoming and Letaief, Khaled B},
	journal = {IEEE Journal on Selected Areas in Communications},
	number = {9},
	pages = {2584--2597},
	publisher = {IEEE},
	title = {Task-Oriented Multi-User Semantic Communications},
	volume = {40},
	year = {2022}}

@inproceedings{mohammed2020distributed,
	author = {Mohammed, Thaha and Joe-Wong, Carlee and Babbar, Rohit and Di Francesco, Mario},
	booktitle = {Proceedings of IEEE Conference on Computer Communications (INFOCOM)},
	organization = {IEEE},
	pages = {854--863},
	title = {{Distributed Inference Acceleration with Adaptive DNN Partitioning and Offloading}},
	year = {2020}}

@article{tishby2000information,
	author = {Tishby, Naftali and Pereira, Fernando C and Bialek, William},
	journal = {arXiv preprint physics/0004057},
	title = {The information bottleneck method},
	year = {2000}}

@article{kang2017neurosurgeon,
	author = {Kang, Yiping and Hauswald, Johann and Gao, Cao and Rovinski, Austin and Mudge, Trevor and Mars, Jason and Tang, Lingjia},
	journal = {ACM SIGARCH Computer Architecture News},
	number = {1},
	pages = {615--629},
	publisher = {ACM New York, NY, USA},
	title = {{Neurosurgeon: Collaborative Intelligence Between the Cloud and Mobile Edge}},
	volume = {45},
	year = {2017}}

@book{goldsmith_2005,
	author = {Goldsmith, Andrea},
	editor = {Cambridge},
	publisher = {Cambridge Univ. Press},
	title = {{Wireless Communications}},
	year = {2005}}

@article{gallagher2018cybersickness,
	author = {Gallagher, Maria and Ferr{\`e}, Elisa Raffaella},
	journal = {Multisensory research},
	number = {7},
	pages = {645--674},
	publisher = {Brill},
	title = {{Cybersickness: a Multisensory Integration Perspective}},
	volume = {31},
	year = {2018}}

@inproceedings{matsubara2022bottlefit,
	author = {Matsubara, Yoshitomo and Callegaro, Davide and Singh, Sameer and Levorato, Marco and Restuccia, Francesco},
	booktitle = {Proceedings of IEEE International Symposium on a World of Wireless, Mobile and Multimedia Networks (WoWMoM)},
	html = {https://arxiv.org/abs/2201.02693},
	title = {{BottleFit: Learning Compressed Representations in Deep Neural Networks for Effective and Efficient Split Computing}},
	year = {2022}}

\vspace{-0.1cm}
\section*{Appendix}
\appendix

\section{Variational Upper Bound and Loss Function Computation}\label{subsec:variational_bound}

In this section, we derive an approximation of the loss function in Equation~\eqref{eqn:loss_ib} that can be effectively implemented to jointly train the \FW modules. This is required as computing the distributions $p(z)$ and $p(y| \hat{z}, s)$ for high-dimensional data with arbitrary distributions is computationally prohibitive. 

\smallskip
\noindent \textbf{Variational Upper Bound Reformulation.}~We define two variational distributions $q(z)$ and $q(y| \hat{z}, s)$ to approximate $p(z)$ and $p(y| \hat{z}, s)$. Specifically, we take the postcoder together with the receiver processing block to be the variational approximation of $p(y| \hat{z}, s)$. The type of this conditional probability $p(y| \hat{z}, s)$ depends on the task. For example, for classification tasks, it has a categorical distribution, and its parameters are found using the postcoder together with the receiver processing.
Using this approximation, the variational upper bound of the first term in Equation~\eqref{eqn:loss_ib} is obtained as\vspace{-0.1cm}
\begin{multline}
    \nonumber
    \mathbb{E}_{p(x, y)} \Bigl\{\mathbb{E}_{p(\hat{z}| x, s)} \bigl[-\log p(y| \hat{z}, s)\bigr]\Bigr\} =\\
    \mathbb{E}_{p(x, y)} \Bigl\{\mathbb{E}_{p(\hat{z}| x, s)} \bigl[-\log q(y| \hat{z}, s)\bigr]\Bigr\}\\
    - \underbrace{\mathbb{E}_{p(\hat{z})}\Bigl\{\mathbb{E}_{p(y| \hat{z}, s)}\bigl[\log \frac{p(y| \hat{z}, s)}{q(y| \hat{z}, s)}\bigr]\Bigr\}}_{D_{KL}\bigl(p(y| \hat{z}, s)|| q(y| \hat{z}, s)\bigr) \geq 0}\\
    \leq \mathbb{E}_{p(x, y)} \Bigl\{\mathbb{E}_{p(\hat{z}| x, s)} \bigl[-\log q(y| \hat{z}, s)\bigr]\Bigr\}.
\end{multline}

Similarly, we compute the upper bound for the second term in Equation~\eqref{eqn:loss_ib} as
\begin{align*}
    D_{KL}\bigl(p(z| x, s) || p(z)\bigr) =&~
    D_{KL}\bigl(p(z| x, s) || q(z)\bigr)\\
    &- \underbrace{D_{KL}\bigl(p(z) || q(z)\bigr)}_{\geq 0}\\
    \leq&~ D_{KL}\bigl(p(z| x, s) || q(z)\bigr).
\end{align*}
Since minimizing this upper bound minimizes the \gls{kl} divergence between $p(z| x, s)$ and $q(z)$, a certain variational prior $q(z)$ can be imposed to induce that prior distribution on the latent symbols. Therefore, to regularize (sparsify) the latent space during training and prevent \FW from memorizing an exact mapping, we use a standard complex normal prior distribution i.e., $q(z) \sim \mathcal{CN}(\mathbf{0}, \mathbf{I})$.

In conclusion, the variational upper bound of the loss function in Equation~\eqref{eqn:loss_ib} writes as
\begin{equation}
\begin{split}
    \label{eqn:loss_vib}
    \mathcal{L}_{VIB} = &~\mathbb{E}_{p(x, y)} \biggl\{\mathbb{E}_{p(s)} \Bigl\{\mathbb{E}_{p(\hat{z}| x, s)} \bigl[-\log q(y| \hat{z}, s)\bigr]\\
    &+ \beta \cdot D_{KL}\bigl(p(z| x, s) || \mathcal{CN}(\mathbf{0}, \mathbf{I})\bigr)\Bigr\}\biggr\}.
\end{split}
\end{equation}
As $p(z| x, s)$ is assumed to be Gaussian and the \gls{kl} divergence between two complex Gaussian distributions has a closed-form expression, the second term in Equation~\eqref{eqn:loss_vib} can be computed analytically.

\smallskip
\noindent \textbf{Monte Carlo Sampling.}~Monte Carlo sampling can be used to obtain an unbiased estimate of the expected values in Equation~\eqref{eqn:loss_vib}. \FW is trained using a gradient descent algorithm on the estimated loss. Note that the precoder output is a distribution, and it should be sampled to obtain a realization. However, such sampling is not differentiable. Therefore, we use the reparameterization trick \cite{kingma2013auto} to make \FW trainable through loss backpropagation. More in detail, given a mini-batch of $N_{mb} \times N_{comp}$ data points $\{(x_i, y_i)\}_{i = 1}^{N_{mb} \times N_{comp}}$, where $N_{mb}$ is the batch size and $N_{comp}$ is the number of multiplexed task inputs, we feed $N_{comp}$ inputs $x$ to the transmitter processing plus the precoder and the Gaussian distribution parameters are obtained for latent symbols in each packet and subcarrier $z_{p, k}$. Then, $N_{mc}$ realizations of latent symbols $z$, where $N_{mc}$ is the number of Monte Carlo samples, are sampled using these distributions. Also, $N_{mc}$ realizations of the \gls{cfr} and $N_{mc}$ realizations of the channel noise $n^f$ are sampled to define $N_{mc}$ realizations of the \gls{csi} ($s$). Hence, the \FW loss function is estimated as follows
\begin{equation}
    \begin{split}
    \label{eqn:loss_mc}
    &\mathcal{\Tilde{L}}_{VIB} = \frac{1}{N_{mb}} \sum_{n_{mb} = 1}^{N_{mb}} \Bigl\{ \frac{1}{N_{mc}} \sum_{n_{mc} = 1}^{N_{mc}} \bigl[- \log q(y_{n_{mb}}| \hat{z}_{n_{mb}, n_{mc}})\\
    &+ \beta \cdot \sum_{p = 1}^{P} \sum_{k = 1}^{K} D_{KL}(p(z_{n_{mb}, n_{mc}, p, k}| x_{n_{mb}}, s_{n_{mc}}) || \mathcal{CN}(\mathbf{0}, \mathbf{I})) \bigr] \Bigr\},
    \end{split}
\end{equation}
where $\hat{z}_{n_{mb}, n_{mc}} = H_{n_{mc}} z_{n_{mb}, n_{mc}} + n_{mc}^f$.

\section{Accuracy on CIFAR-100 and SVHN}\label{subsec:results_other}

We report the image classification accuracy using the CIFAR-100 and SVHN datasets in Tables~\ref{tab:acc_CIFAR100}-\ref{tab:acc_SVHN}. We compare the performance of \FW with the other baselines for all the three \gls{dnn} models considered for this task. The results are in line with the discussion in Section~\ref{subsec:task_accuracy}: \FW lead to a slight degradation in performance with respect to the full protocol stack approaches while surpassing all \gls{phy} strategies. Moreover, by comparing the results in a cross-dataset fashion, we can see that the multiplexing performance depends on the task complexity, e.g, recognizing images from 10 classes (CIFAR-10, SVHN) or from 100 classes (CIFAR-100), as discussed in Section~\ref{subsec:scalability_evaluation}. We remind that the performance of full-stack approaches is not affected by the communication scenario.\vspace{-0.2cm}  

\begin{table}[h]
  \small  
  \caption{Image classification. Experimental evaluation of \FW and baselines for different \glspl{dnn} on CIFAR-100.\vspace{-0.3cm}}
  \label{tab:acc_CIFAR100}
  \begingroup
  \setlength{\tabcolsep}{2pt}
  \renewcommand{\arraystretch}{0.85}  
  \begin{tabular}{@{}c|cccccc@{}}
    \toprule
    \textbf{WRN-28-10} & MNet & BF-9 & BF-3 & MCR2 & PhyDNN & \textbf{SM}\\
    \midrule
    \gls{los} & \multirow{2}{*}{79.20 \%} & \multirow{2}{*}{81.32 \%} & \multirow{2}{*}{80.63 \%} & 74.68 \% & 75.75 \% & 78.87 \%\\
    \gls{nlos} & & & & 71.46 \% & 72.52 \% & 76.62 \%\\
    \midrule
    \textbf{WRN-16-8} & MNet & BF-9 & BF-3 & MCR2 & PhyDNN & \textbf{SM}\\
    \midrule
    \gls{los} & \multirow{2}{*}{76.84 \%} & \multirow{2}{*}{77.58 \%} & \multirow{2}{*}{77.12 \%} & 71.87 \% & 72.74 \% & 75.13 \%\\
    \gls{nlos} & & & & 68.14 \% & 69.30 \% & 72.31 \%\\
    \midrule
    \textbf{WRN-10-4} & MNet & BF-9 & BF-3 & MCR2 & PhyDNN & \textbf{SM}\\
    \midrule
    \gls{los} & \multirow{2}{*}{66.23 \%} & \multirow{2}{*}{69.08 \%} & \multirow{2}{*}{67.38 \%} & 60.55 \% & 61.43 \% & 63.58 \%\\
    \gls{nlos} & & & & 58.42 \% & 59.88 \% & 61.14 \%\\
    \bottomrule
  \end{tabular}
  \endgroup
\end{table}

\vspace{0.2cm}
\begin{table}[h]
  \small  
  \caption{Image classification. Experimental evaluation of \FW and baselines for different \glspl{dnn} on SVHN.\vspace{-0.3cm}}
  \label{tab:acc_SVHN}
  \begingroup
  \setlength{\tabcolsep}{2pt}
  \renewcommand{\arraystretch}{0.85}  
  \begin{tabular}{@{}c|cccccc@{}}
    \toprule
    \textbf{WRN-28-10} & MNet & BF-9 & BF-3 & MCR2 & PhyDNN & \textbf{SM}\\
    \midrule
    \gls{los} & \multirow{2}{*}{96.88 \%} & \multirow{2}{*}{97.11 \%} & \multirow{2}{*}{97.08 \%} & 92.48 \% & 94.92 \% & 96.51 \%\\
    \gls{nlos} & & & & 90.02 \% & 92.68 \% & 94.73 \%\\
    \midrule
    \textbf{WRN-16-8} & MNet & BF-9 & BF-3 & MCR2 & PhyDNN & \textbf{SM}\\
    \midrule
    \gls{los} & \multirow{2}{*}{96.69 \%} & \multirow{2}{*}{96.84 \%} & \multirow{2}{*}{96.67 \%} & 92.45 \% & 95.27 \% & 96.32 \%\\
    \gls{nlos} & & & & 89.40 \% & 92.97 \% & 94.91 \%\\
    \midrule
    \textbf{WRN-10-4} & MNet & BF-9 & BF-3 & MCR2 & PhyDNN & \textbf{SM}\\
    \midrule
    \gls{los} & \multirow{2}{*}{94.33 \%} & \multirow{2}{*}{94.84 \%} & \multirow{2}{*}{94.53 \%} & 91.14 \% & 92.67 \% & 93.61 \%\\
    \gls{nlos} & & & & 89.02 \% & 90.80 \% & 92.12 \%\\
    \bottomrule
  \end{tabular}
  \endgroup
\end{table}
\pagebreak
\clearpage
\end{document}